\documentclass[11pt]{article}
\usepackage{psfig}
\newtheorem{proposition}{Proposition}
\usepackage{fullpage}
\usepackage{amssymb}
\title{Soliton Interactions in Perturbed Nonlinear Schr\"odinger
Equations} \author{James A. Besley\thanks{ Previously affiliated with
the Optical Sciences Centre in the Research School of Physical
Sciences and Engineering at ANU.  Current address: Racal Research
Ltd., Worton Drive, Worton Grange Industrial Estate, Reading,
Berkshire, RG2 0SB, UK.  Email: {\tt
James.Besley@rrl.co.uk}}\\Australian National University \and Peter
D. Miller\thanks{ Address: Department of Mathematics and Statistics,
P. O. Box 28M, Monash University, VIC 3800, Australia.  Email: {\tt
millerpd@mail.maths.monash.edu.au}}\\Institute for Advanced Study
and Monash University \and Nail N. Akhmediev\thanks{Address: Optical
Sciences Centre, Australian National University, Canberra, ACT 0200,
Australia.  Email: {\tt nna124@rsphysse.anu.edu.au}}\\Australian
National University} 
\date{8 June 1999\\Submitted to {\em Phys. Rev. E}\\Revised 10 February 2000
}

\begin{document}
\maketitle
\begin{abstract}
We use multiscale perturbation theory in conjunction with the inverse
scattering transform to study the interaction of a number of solitons
of the cubic nonlinear Schr\"odinger equation under the influence of a
small correction to the nonlinear potential.  We assume that the
solitons are all moving with the same velocity at the initial instant;
this maximizes the effect each soliton has on the others as a
consequence of the perturbation.  Over the long time scales that we
consider, the soliton amplitudes remain fixed, while their
center of mass coordinates obey Newton's equations with a force law
for which we present an integral formula.  For the interaction of two
solitons with a quintic perturbation term we present more details
since symmetries --- one related to the form of the perturbation and
one related to the small number of particles involved --- allow the
problem to be reduced to a one-dimensional one with a single
parameter, an effective mass.  The main results include calculations
of the binding energy and oscillation frequency of nearby solitons in
the stable case when the perturbation is an attractive correction to
the potential and of the asymptotic ``ejection'' velocity in the
unstable case.  Numerical experiments illustrate the accuracy of the
perturbative calculations and indicate their range of validity.
\end{abstract}

\section{Introduction}
This paper is concerned with the asymptotic behavior of solutions of
the initial-value problem for the perturbed nonlinear Schr\"odinger
equation (NLSE)
\begin{equation}
i\partial_t
\psi +\frac{1}{2}\partial_x^2\psi + |\psi|^2\psi + p[\psi,\psi^*]=0\,,
\label{eq:fundamentalIVP}
\end{equation}
subject to the initial condition $\psi(x,0)=\psi_0(x)$ for certain
initial fields $\psi_0(x)$, in the limit when the perturbation term
$p[\psi,\psi^*]$ becomes formally small.  The unperturbed problem,
when $p[\psi,\psi^*]\equiv 0$ in (\ref{eq:fundamentalIVP}), is
well-known to be solvable \cite{Zakharov:1972} by an inverse
scattering transform, one consequence of which is the
existence of finite energy soliton solutions that are dynamically
stable and robust with respect to collisions.  The unperturbed NLSE
arises in two different physical situations in modern optics
\cite{Akhmediev:1997}.  Firstly, for high-speed
telecommunications, (\ref{eq:fundamentalIVP}) describes
the propagation of light wave-packets along an optical fiber.  In this
interpretation, $t$ is the spatial coordinate along the fiber and $x$
is the retarded time variable for the signal; accordingly the solitons
of the unperturbed problem (and usually also the solitary waves of the
perturbed problem, when they exist) are called temporal solitons.  The
suggestion by Hasegawa and Tappert in 1973
\cite{Hasegawa:1973,Hasegawa:1973b} that temporal solitons, being
immune to dispersion, might serve as bits in a high-speed data stream
has since generated a large body of work, much of which
is comprehensively reviewed in \cite{Agrawal:1989,Hasegawa:1995}.
Secondly, for photonic switching devices, 
(\ref{eq:fundamentalIVP}) describes the stationary envelope of
monochromatic light waves in a planar waveguide under the paraxial
approximation.  Here $x$ and $t$ are both spatial variables, with $t$
being the propagation direction and $x$ being the transverse
direction; accordingly the solitons of the unperturbed problem are
called spatial solitons.

In both of these applications, the cubic term $|\psi|^2\psi$ in
(\ref{eq:fundamentalIVP}) models the Kerr effect in which the
refractive index of the material depends linearly on the local
intensity of light.  For weakly nonlinear effects, when intensities
are not too large, this effect is dominant in isotropic materials like
glass.  This fact, along with the integrability afforded by neglecting
$p[\psi,\psi^*]$, makes the unperturbed NLSE one of the most important
models in modern optics.

Of course, real materials can have a complicated dependence of
refractive index on intensity, for which the Kerr effect is only an
idealization.  Modeling such phenomena requires introducing
corrections to the coefficient $|\psi|^2$ in the cubic term of the
NLSE.  The perturbative term $p[\psi,\psi^*]$ might also include corrections
related to higher-order dispersion, the Raman effect, self-steepening of
pulses, etc.  
In this paper we will consider only the influence of higher-order
nonlinearity on solitons of the unperturbed NLSE.  For spatial
solitons in photorefractive media, such a perturbation can be the main
factor influencing propagation.  In particular, we take the
perturbation in (\ref{eq:fundamentalIVP}) in the form of a quintic
term
\begin{equation}
p[\psi,\psi^*]=\sigma\epsilon^2|\psi|^4\psi\,,
\label{eq:quintic}
\end{equation}
where $\epsilon>0$ is a small parameter and $\sigma=\pm 1$.

In view of the possibility of using solitons as bits in optical fibers
or dynamically controllable switches in planar waveguides, it is of
some interest to determine the effect of such a perturbation on the
solitons of the unperturbed problem.  If one considers an initial
condition $\psi_0(x)$ that is a ``snapshot'' of a simple soliton
solution of the unperturbed problem, then there are many approaches
available to study the perturbed evolution.  Because the unperturbed
soliton is stationary in some Galilean frame, the main effect of
$p[\psi,\psi^*]$ will be an adiabatic adjustment of the soliton's
amplitude and phase parameters.  This fact, together with the
simplicity of the form of the soliton solution, means that direct
perturbative methods can be used to study their slow evolution.  In
particular, variational methods and multiscale methods applied
directly to (\ref{eq:fundamentalIVP}) often give valid
results.  These perturbative methods are dynamical in origin and
capture effects on finite but long scales.  Other methods can be used
to answer infinite time questions concerning the persistence of
solitary waves.  In fact in the presence of quite general
perturbations solitary waves continue to exist for arbitrary
$\epsilon$ \cite{Karpman:1977,Lamb:1980} and these can be expressed in
closed form in some cases \cite{Akhmediev:1997}.

The presence of more than one soliton complicates the analysis.  If
the solitons are isolated then the field may be approximated as a sum
of solitons plus a small error term, and the adiabatic coupling among
the solitons may be calculated by several methods.  Note that if the
solitons are moving with respect to each other then they will always
be in isolation except possibly for a short time.  An early analysis
of this kind was carried out by Gordon \cite{Gordon:1983}, who studied
the exact two-soliton solution of the unperturbed NLSE 
for equal velocities.  When the solitons are well-separated,
there is an effective force between them (even in the unperturbed
NLSE) that varies sinusoidally with their phase difference.
This phase difference grows linearly in $t$ if the
solitons differ in amplitude.  The force is therefore zero on
average \cite{Desem:1987} and one expects periodic motion.  This is a
physical explanation of the mathematical fact that the intensity
$|\psi|^2$ of the exact two-soliton solution for equal velocities
is a periodic function of $t$.  An extension of this 
argument to perturbed problems was given by Ankiewicz
\cite{Ankiewicz:1995}, who obtained a simple description of soliton
interactions with the use of complex averaged potentials.  Again, the
essential assumption is that the solitons are well-separated in $x$,
so that the field may be approximated as a sum of solitons.  If the
solitons are close to each other, nonlinear interference effects cause
the field to adopt a form very different from the linear superposition
of individual solitons, and therefore a different approach is needed.
Often, one turns to numerics to study the interactions of
solitons in various media (see, for example,
\cite{Desem:1987b,Desem:1987c,Boardman:1995}) without the restriction
of the solitons being isolated.

In the scattering transform domain, where the dynamics of the
unperturbed NLSE are trivial, a state in which two solitons are
close to each other in $x$ has the same spectrum as a state in which
they are far apart.  This suggests that for studying the
influence of perturbations on multisoliton bound states (that is,
several solitons traveling with the same velocity, represented by a
collection of eigenvalues of the Zakharov-Shabat equations
with the same real part) it is best to carry out the analysis in the
transform domain using soliton perturbation theory
\cite{Karpman:1977,KaupNewell:1978,Lamb:1980}.  With
$p[\psi,\psi^*]\neq 0$, the evolution of the scattering data is no longer
trivial, and thus the scope of possible dynamics in near-integrable
systems like (\ref{eq:fundamentalIVP}) is much greater than in the
unperturbed NLSE, including effects like repulsion, attraction, and
energy exchange among bound or colliding solitons.  Other techniques
that have been used to study these effects 
include the judicious use of conserved quantities
\cite{Akhmediev:1997}, variational methods
\cite{Anderson:1983,Whitham:1974}, ``equivalent particle'' approaches
\cite{Aceves:1989,Aceves:1990}, and of course, numerics.

In this paper, we use soliton perturbation theory to study
perturbations of the nonlinear potential in (\ref{eq:fundamentalIVP}),
for initial conditions $\psi_0(x)$ that are snapshots of multisoliton
bound states of the unperturbed NLSE.  With respect to treating the
solitons in isolation, this is a worst-case scenario since
in the unperturbed NLSE a tightly-bound state of solitons will
remain so for all time.  Nonetheless, it is a scenario of some
interest, in particular for the quintic perturbation
(\ref{eq:quintic}).  If $\sigma=+1$, then it is known that the
solution remains bound, and this case has been studied using
conservation laws \cite{Buryak:1994}.  If $\sigma=-1$, then the bound
state becomes destabilized.  Recently it was shown \cite{Artigas:1997}
by simulations of (\ref{eq:fundamentalIVP}) that the
instability causes the bound state to divide into isolated solitons
that are ejected from the origin with nonzero relative
velocities.  On the time scales over which this splitting occurs, the
solitons do not appear to exchange energy.  In mathematical terms, 
each eigenvalue in the bound state ensemble,
originally confined to the imaginary axis (zero
velocity), appears to slowly ``grow'' a real part while its imaginary part
remains fixed.  Once the solitons escape, they no longer interact and
the velocities no longer change.  The wave guidance properties of Y-junctions
engineered from such splittings of spatial solitons have also been 
analyzed \cite{OQE}.

By considering the relative velocities to be small, we will find an
integral formula that expresses the asymptotic velocity difference
between a pair of initially co-propagating solitons destabilized by
the quintic perturbation (\ref{eq:quintic}) with $\sigma=-1$.  
Along the way, we will write down a coupled system of differential
equations that describes the interaction of any number of solitons
under more general perturbations over long time scales.  These
equations are just Newton's equations for a system of interacting
particles in one space dimension; the particle coordinates have the
interpretation of the soliton centers of mass.  The force is
translationally invariant, conserves the total momentum, and 
is also proportional to $\sigma$, so the forces giving rise to
attraction and repulsion are related just by a
change of sign.  For the interaction of two solitons, 
the problem may be reduced to a single degree of
freedom, the relative separation of the solitons.  The force law
scales simply with the (fixed) amplitudes, which have the
interpretation of masses.  The result is a one-parameter family of
problems indexed by a normalized effective mass.  If the
separation is small in the attractive case $\sigma=+1$, the
force is nearly linear and the frequency of motion becomes a function
of the normalized effective mass.  We calculate this frequency, a
quantity that is connected with the vibrations of solitons that are
infinitely close, a limit opposite to the well-separated case.

Our paper begins in \S\ref{sec:exact} with a review of the theory of
the scattering transform for the Zakharov-Shabat eigenvalue problem
and of the inverse theory that holds in the reflectionless case.  We
also recall the derivation of the exact equations of motion in the
transform domain corresponding to the perturbed NLSE
(\ref{eq:fundamentalIVP}).  Then, in \S\ref{sec:perturb} we consider
perturbations of the form $p[\psi,\psi^*]=\epsilon^2 W(|\psi|^2)\psi$
and apply multiscale perturbation theory to find asymptotic solutions
of the equations of motion in the transform domain.  The
approximations are uniformly valid as $\epsilon\downarrow 0$ on
expanding time intervals of length $\epsilon^{-1}$, and are given in
terms of solutions of Newton's equations for particles interacting in
one dimension under a force law that has several universal features.
In \S\ref{sec:two} we focus on the quintic perturbation
(\ref{eq:quintic}) and study the interaction of two solitons.  We
reduce the problem to the motion of a single particle and then
explicitly perform the averaging required to remove secular terms from
the asymptotic expansion.  This leaves the force law in the form of a
1D integral that we study numerically.  We use it to
compute the ``ejection'' velocity observed by Artigas
et. al. \cite{Artigas:1997} in the unstable case and the harmonic
frequency of tightly-bound solitons in the stable case.  Finally, we
compare the results of perturbation theory with direct simulations of
(\ref{eq:fundamentalIVP}).  The Appendix contains the more
cumbersome formulae that nonetheless are among our main analytical
results.

Regarding notation, we will use stars for complex conjugation, and matrices
will be written with bold letters, except for the Pauli matrices
\begin{equation}
\sigma_1:= \left[\begin{array}{cc} 0 & 1 \\ 1 & 0\end{array}\right]\,,
\hspace{0.3 in}
\sigma_2:= \left[\begin{array}{cc}0 & -i \\ i & 0\end{array}\right]\,,
\hspace{0.3 in}
\sigma_3:= \left[\begin{array}{cc}1 & 0 \\ 0 & -1\end{array}\right]\,.
\end{equation}

\section{Exact Inverse Scattering Theory For The Perturbed NLSE}
\label{sec:exact}
Here, we review the known inverse scattering theory for the
Zakharov-Shabat eigenvalue problem to fix our notation.  In general,
we wish to consider (\ref{eq:fundamentalIVP}) where $p[\psi,\psi^*]$
is a polynomial in $\psi$, $\psi^*$, and their $x$-derivatives.  The
field $\psi$ is taken to be in the Schwartz space as a function of
$x$.

\subsection{Scattering data.}
We will work with the scattering transform of $\psi$, a map that
associates to the complex field $\psi$ at each fixed time a set of
``scattering data'' from which $\psi$ can be reconstructed by
inverting the map.  As is well-known, the advantage of this is that
the time evolution of the scattering data corresponding to the time
evolution of $\psi$ is trivial when $p\equiv 0$.  Consequently, when
$|p|\ll 1$, this proves to be a useful setting for perturbation
theory.

Fix $t$, and assume the complex function $\psi(x,t)$
to be given.  For $\lambda\in{\mathbb R}$ denote by ${\bf
M}^\pm(x,t,\lambda)$ the $2\times 2$ matrix solutions of the linear
differential equation
\begin{equation}
\partial_x{\bf M}^\pm={\bf L}{\bf M}^\pm:=
\left[\begin{array}{cc}-i\lambda & \psi\\-\psi^* & i\lambda
\end{array}\right]{\bf M}^\pm\,,
\label{eq:linsys}
\end{equation}
satisfying the boundary conditions $ {\bf
M}^\pm(x,t,\lambda)\exp(i\lambda\sigma_3 x)\rightarrow {\mathbb I}$ as
$x\rightarrow \pm\infty$ Since ${\bf L}$ is traceless, these boundary
conditions guarantee that these matrices are unimodular for all $x$.
For each $\lambda$ there can only be two linearly independent column
vector solutions of (\ref{eq:linsys}); therefore there is a matrix
${\bf S}(t,\lambda)$, $\lambda\in{\mathbb R}$, the {\em scattering
matrix}, such that
\begin{equation}
{\bf M}^-(x,t,\lambda)={\bf M}^+(x,t,\lambda){\bf S}(t,\lambda)\,.
\label{eq:scatdef}
\end{equation}
The first column of ${\bf M}^-(x,t,\lambda)$ and the second column of
${\bf M}^+(x,t,\lambda)$ turn out to be boundary values of
analytic functions for $\Im(\lambda)>0$, while the second column of
${\bf M}^-(x,t,\lambda)$ and the first column of ${\bf
M}^-(x,t,\lambda)$ are the boundary values of analytic functions for
$\Im(\lambda)<0$.  Adjoining the second column of ${\bf
M}^+(x,t,\lambda)$ on the right of the first column of
(\ref{eq:scatdef}) and taking determinants gives
\begin{equation}
S_{11}(t,\lambda)=\det(M_1^-(x,t,\lambda),M_2^+(x,t,\lambda))\,,
\label{eq:DET+}
\end{equation}
which is therefore the boundary value of a function analytic for 
$\Im(\lambda)>0$.
Likewise
$
S_{22}(t,\lambda)=\det(M_1^+(x,t,\lambda),M_2^-(x,t,\lambda))
$
is the boundary value of a function analytic for 
$\Im(\lambda)<0$.

Fix $\lambda\in{\mathbb R}$.  Then, from (\ref{eq:linsys}),
$ {\bf M}^{\pm *}=\sigma_2{\bf M}^\pm\sigma_2 $, and thus $ {\bf
S}^*=\sigma_2{\bf S}\sigma_2$, so that $S_{22}=S_{11}^*$ and
$S_{21}=-S_{12}^*$.  In particular, this means that as an analytic
function for $\Im(\lambda)<0$, $
S_{22}(t,\lambda)=S_{11}(t,\lambda^*)^* $.  Also, for
$\lambda\in{\mathbb R}$ the fact that $\det ({\bf S})=1$ implies the
normalization condition $ |S_{11}|^2 + |S_{12}|^2 = 1 $.

The analytic function $S_{11}(t,\lambda)$ with $\Im(\lambda)>0$ may
have zeros $\lambda_1(t),\dots, \lambda_N(t)$.
The determinant formula (\ref{eq:DET+})
then shows that there exist complex numbers
$\gamma_1(t),\dots,\gamma_N(t)$ such that
\begin{equation}
M_2^+(x,t,\lambda_k(t))=\gamma_k(t)M_1^-(x,t,\lambda_k(t))\,,\hspace{0.3 in}
k=1,\dots,N\,.
\end{equation}
The conjugation symmetry of ${\bf M}^\pm(x,t,\lambda)$ for
$\lambda\in{\mathbb R}$, when extended to the complex plane, implies
that at the complex conjugate points $\lambda_k(t)^*$ where
$S_{22}(t,\lambda)$ vanishes,
$M_1^+(x,t,\lambda_k(t)^*)=-\gamma_k(t)^*M_2^-(x,t,\lambda_k(t)^*)$,
for $k=1,\dots,N$.  Since $S_{11}(t,\lambda)\rightarrow 1$ as
$\lambda\rightarrow\infty$ with $\Im(\lambda)>0$, Hilbert
transform theory can be used in conjunction with the normalization
condition to express $S_{11}(t,\lambda)$ for $\Im(\lambda)>0$ in terms
of its zeros and the values of $S_{12}(t,\lambda)$ on the real axis
\cite{fandt}:
\begin{equation}
S_{11}(t,\lambda)=\left(\prod_{k=1}^N\frac{\lambda-\lambda_k(t)}
{\lambda-\lambda_k(t)^*}\right)\exp\left(\frac{1}{2\pi i}
\int_{-\infty}^\infty\frac{\log(1-|S_{12}(t,\mu)|^2)}{\mu-\lambda}\,d\mu\right)
\,.
\label{eq:HilbertXF}
\end{equation}

The so-called ``trace formulae'' that equate certain functionals of
the potential $\psi $ to functionals of the scattering data will
be useful below.  In particular, we will use the
formula:
\begin{equation}
P[\psi,\psi^*]:=\int_{-\infty}^\infty\Im
(\psi\partial_x\psi^*)\,dx=
\frac{1}{2\pi}\int_{-\infty}^\infty\mu\log(1-|S_{12}(t,\mu)|^2)\,d\mu-
\sum_{k=1}^N\Im(\lambda_k(t)^2)\,.
\label{eq:trace}
\end{equation}
This functional (not to be confused with the 
perturbation $p[\psi,\psi^*]$) has the interpretation of the total
momentum of the wave function $\psi(x,t)$.  For the unperturbed
problem, as well as in the presence of many physically important
perturbations, the total
momentum is a constant of motion.

\subsection{Reconstruction of the potential in the reflectionless case.}
\label{sec:reflectionless}
The miracle of inverse scattering theory is that for each fixed $t$,
the potential $\psi(x,t)$ can be recovered from its scattering data,
namely the reflection coefficient $S_{12}(t,\lambda)$ for
$\lambda\in{\mathbb R}$, the eigenvalues
$\{\lambda_k(t)\}$ with $\Im(\lambda_k)>0$, and the
proportionality constants $\{\gamma_k(t)\}$.  The
reconstruction is particularly simple if $S_{12}(t,\lambda)\equiv 0$
as a function of $\lambda$ for some $t$, since it then follows
from (\ref{eq:HilbertXF}) that
\begin{equation}
S_{11}(t,\lambda)=\prod_{k=1}^N\frac{\lambda-\lambda_k(t)}
{\lambda-\lambda_k(t)^*}\,,
\end{equation}
which extends to $\Im(\lambda)<0$ as a meromorphic function.
Similarly one sees that $S_{22}(t,\lambda)=1/S_{11}(t,\lambda)$ and
that $S_{21}(t,\lambda)\equiv 0$.  Since ${\bf S}(t,\lambda)$ is
diagonal in this case, the solution matrices ${\bf
M}^\pm(x,t,\lambda)$ can be expressed in terms of a common solution
matrix ${\bf U}(x,t,\lambda)$ by setting
$
{\bf M}^\pm(x,t,\lambda):={\bf U}(x,t,\lambda){\bf N}^\pm(t,\lambda)
$,
where \cite{BesleyThesis:1998}
\begin{equation}
{\bf N}^\pm(t,\lambda)=
\sigma_1^{\frac{1\mp 1}{2}}
{\rm diag\,}\left(\prod_{k=1}^N(\lambda-\lambda_k(t))^{-1},
\prod_{k=1}^N(\lambda-\lambda_k(t)^*)^{-1}\right)\sigma_1^{\frac{1\mp 1}{2}}\,.
\end{equation}
The columns of ${\bf
U}(x,t,\lambda)=(U_1(x,t,\lambda),U_2(x,t,\lambda))$ necessarily satisfy the
relations
\begin{equation}
U_2(x,t,\lambda_k(t))=\gamma_k(t)U_1(x,t,\lambda_k(t))\,,\hspace{0.3 in}
-\gamma_k(t)^*U_2(x,t,\lambda_k(t)^*)=U_1(x,t,\lambda_k(t)^*)\,,
\label{eq:relations}
\end{equation}
for all $k=1,\dots,N$.  It follows that ${\bf U}(x,t,\lambda)$
takes the simple form
\begin{equation}
{\bf U}(x,t,\lambda)=\left(\lambda^N{\mathbb I}+\sum_{p=0}^{N-1}\lambda^p
{\bf U}^{(p)}(x,t)\right)\exp(-i\lambda\sigma_3 x)\,,
\end{equation}
that is, a polynomial in $\lambda$ times an exponential, where the
matrix coefficients ${\bf U}^{(p)}(x,t)$ are determined uniquely from
(\ref{eq:relations}).  This means that
(\ref{eq:relations}) can be viewed as a linear algebraic 
system of $4N$ equations in $4N$ unknowns, the matrix elements of 
${\bf U}^{(p)}(x,t)$.  Moreover, it can be shown that ${\bf
U}$ constructed in this way is satisfies $\partial_x{\bf
U}={\bf LU}$ if and only if the potential function in
${\bf L}$ is
\begin{equation}
\psi(x,t)=2iU^{(N-1)}_{12}(x,t)\,.
\end{equation}
This formula reconstructs $\psi(x,t)$ from the discrete scattering
data $\{\lambda_k(t)\}$ and $\{\gamma_k(t)\}$ in the
``reflectionless'' case when $S_{12}(t,\lambda)\equiv 0$.  This
treatment of multisoliton potentials via the matrix $\bf U$ follows
Krichever \cite{Krichever:1977}, Manin \cite{Manin:1979}, and
Date \cite{Date:1982}.  See \cite{Miller:1996} for a relevant application.

\subsection{Dynamics of the scattering data.}
We now recall how the data evolve in $t$ when
$\psi$ satisfies (\ref{eq:fundamentalIVP}).
The motivating observation \cite{Lamb:1980} is that
(\ref{eq:fundamentalIVP}) can be cast in matrix form:
\begin{equation}
i\partial_t{\bf L}-\partial_x{\bf B}+[{\bf L},{\bf B}]+{\bf P}={\bf 0}\,,
\label{eq:matrixform}
\end{equation}
where the matrix ${\bf L}$ is the one appearing in the linear scattering
problem (\ref{eq:linsys}), and where
\begin{equation}
{\bf B}=\left[\begin{array}{cc}
\displaystyle \lambda^2 - \frac{1}{2}|\psi|^2 &
\displaystyle i\lambda\psi-\frac{1}{2}\partial_x\psi\\\\
\displaystyle -i\lambda\psi^*-\frac{1}{2}\partial_x\psi^* &
\displaystyle -\lambda^2 +\frac{1}{2}|\psi|^2
\end{array}\right]\,,\hspace{0.3 in}
{\bf P}=\left[\begin{array}{cc}0 & p[\psi,\psi^*] \\\\
p[\psi,\psi^*]^* & 0\end{array}\right]\,.
\end{equation}
Using the fact that ${\bf M}^\pm$ satisfies 
(\ref{eq:linsys}),
multiply (\ref{eq:matrixform}) on the right by ${\bf
M}^\pm$ and find
\begin{equation}
(\partial_x-{\bf L})(i\partial_t-{\bf B}){\bf M}^\pm+
{\bf P}{\bf M}^\pm={\bf 0}\,.
\end{equation}
This equation is solved for $(i\partial_t-{\bf B}){\bf
M}^\pm$ by variation of parameters.  Introducing a new
unknown ${\bf J}^\pm(x,t,\lambda)$ defined through
the relation $ (i\partial_t-{\bf
B}){\bf M}^\pm={\bf M}^\pm {\bf J}^\pm $, one finds that ${\bf J}^\pm$
satisfies $ \partial_x{\bf J}^\pm= -{\bf M}^{\pm -1}{\bf P}{\bf
M}^\pm$.  We now integrate to find ${\bf J}^\pm$ explicitly, taking into
account the boundary conditions satisfied by ${\bf
M}^\pm$ as $x\rightarrow\pm\infty$ and the fact that
in both limits ${\bf B}\rightarrow\lambda^2 \sigma_3$.
With the use of these explicit formulae for ${\bf
J}^\pm$ the equations $ (i\partial_t-{\bf B}){\bf
M}^\pm={\bf M}^\pm {\bf J}^\pm $ become equations of motion for the
matrices ${\bf M}^\pm$:
\begin{equation}
(i\partial_t-{\bf B}){\bf M}^\pm=\displaystyle
{\bf M}^\pm
\left(-\lambda^2\sigma_3+\int_x^{\pm\infty}{\bf M}^{\pm -1}
{\bf P}{\bf M}^\pm\,dx'\right)\,.
\label{eq:EquationsForMatrices}
\end{equation}

As written, (\ref{eq:EquationsForMatrices}) does not make sense for
$\Im(\lambda)\neq 0$.  But for $\lambda\in{\mathbb R}$, the columns
$M^-_1$ and $M^+_2$ are the boundary values of functions analytic for
$\Im(\lambda)>0$, and we will also need equations for them that hold
for $\Im(\lambda)>0$.  To this end, we introduce the matrix $ {\bf
M}(x,t,\lambda):=(M^-_1,M^+_2)$, and as before
define the new unknown ${\bf J}(x,t,\lambda)=(J_1,J_2)$ through
the relation $
(i\partial_t-{\bf B}){\bf M}={\bf M} {\bf J} $, and then 
integrate:
\begin{equation}
J_1=
\left[\begin{array}{c}-\lambda^2\\0\end{array}\right]-
\int_{-\infty}^x{\bf M}^{-1}{\bf P}M^-_1\,dx'\,,
\hspace{0.3 in}
J_2=
\left[\begin{array}{c}0\\\lambda^2\end{array}\right]+
\int_x^\infty{\bf M}^{-1}{\bf P}M^+_2\,dx'\,.
\end{equation}
As before, these expressions are used in $ (i\partial_t-{\bf B}){\bf
M}={\bf M} {\bf J}$ to yield the equation of motion for ${\bf
M}$, valid for $\Im(\lambda)>0$ {\em except}
at $\{\lambda_k\}$, where ${\bf M}$ fails to be invertible.  Each
singularity is, however, removable, since $\det{\bf M}=S_{11}$ and
hence (writing $M^{\pm\prime}_{jk}$ for $M_{jk}^\pm(x',t,\lambda)$)
\begin{equation}
{\bf M}(x,t,\lambda){\bf M}(x',t,\lambda)^{-1}=\frac{1}{S_{11}}
\left[\begin{array}{cc}
M^-_{11}M^{+\prime}_{22}-M^+_{12}M^{-\prime}_{21} & 
M^+_{12}M^{-\prime}_{11}-M^-_{11}
M^{+\prime}_{12}\\
M^-_{21}M^{+\prime}_{22}-M^{-\prime}_{21}M^+_{22} & 
M^{-\prime}_{11}M^+_{22}-M^{+\prime}_{12}
M^-_{21}\end{array}\right]\,.
\end{equation}
We make the natural assumption that the (isolated) zeros
$\lambda=\lambda_k(t)$ of the denominator $S_{11}(t,\lambda)$ are
simple \cite{fandt}.
But then the numerator of each entry is analytic
at $\lambda=\lambda_k(t)$ and is easily seen to vanish there, thus
cancelling the singularity.  Hence, the evolution equation for ${\bf
M}$ makes sense as $\lambda\rightarrow\lambda_k(t)$.  We
accordingly introduce the notation
\begin{equation}
{\bf H}_k(x,x',t):= \lim_{\lambda\rightarrow\lambda_k(t)}
{\bf M}(x,t,\lambda){\bf M}(x',t,\lambda)^{-1}\,.
\end{equation}

The equations of motion for ${\bf M}^\pm$ and ${\bf M}$
determine the evolution of the scattering data.  Using ${\bf S}:={\bf
M}^{+-1}{\bf M}^-$, for real $\lambda$ one finds 
\begin{equation}
i\partial_t{\bf S}=
-{\bf M}^{+-1}i\partial_t{\bf M}^+\cdot
{\bf M}^{+-1}{\bf M}^-+
{\bf M}^{+-1}i\partial_t{\bf M}^-
=
-{\bf M}^{+-1}i\partial_t{\bf M}^+\cdot
{\bf S}+{\bf M}^{+-1}i\partial_t{\bf M}^-\,.
\end{equation}
Substituting from 
(\ref{eq:EquationsForMatrices}) yields
\begin{equation}
i\partial_t{\bf S}=\lambda^2\sigma_3{\bf S}-\int_x^\infty{\bf M}^{+\,\,-1}
{\bf P}{\bf M}^+\,dx'\cdot{\bf S}-\lambda^2{\bf S}\sigma_3-
{\bf S}\int_{-\infty}^x{\bf M}^{-\,\,-1}{\bf P}{\bf M}^-\,dx'\,.
\end{equation}
Finally, since ${\bf S}$ does not depend on $x$, it may be brought
inside the integrals.  With the use of its definition the integrals
are combined, giving the equation of motion:
\begin{equation}
i\partial_t{\bf S}(t,\lambda)+\lambda^2[{\bf S}(t,\lambda),\sigma_3]+
\int_{-\infty}^\infty {\bf M}^+(x',t,\lambda)^{-1}{\bf P}{\bf M}^-(x',t,\lambda)\,dx'={\bf 0}\,.
\end{equation}
Note that since 
$\bf P$ is off-diagonal, the equation for
$S_{11}(t,\lambda)$ only involves quantities analytic
for $\Im(\lambda)>0$.  Likewise, the equation for
$S_{22}(t,\lambda)$ only involves quantities analytic
for $\Im(\lambda)<0$.

The equation of motion for the reflection coefficient
$S_{12}(t,\lambda)$ is contained in that for $\bf S$:
\begin{equation}
i\partial_tS_{12}(t,\lambda)-2\lambda^2S_{12}(t,\lambda)+
\int_{-\infty}^\infty \left[{\bf M}^+(x',t,\lambda)^{-1}{\bf P}
{\bf M}^-(x',t,\lambda)\right]_{12}dx'=0\,.
\label{eq:reflectionevolve}
\end{equation}
The integrand here is $ p[\psi,\psi^*]M^+_{22}M^-_{22}-
p[\psi,\psi^*]^*M^+_{12}M^-_{12} $, evaluated at $x'$, $t$, and
$\lambda$, which generally only makes sense for $\lambda\in{\mathbb
R}$, as required.  Now, the expression defining the zeros
$\lambda_k(t)$ of $S_{11}(t,\lambda)$ is $ S_{11}(t,\lambda_k(t))=0$.
Differentiating with respect to $t$ gives
\begin{equation}
i\partial_tS_{11}(t,\lambda_k(t))+i\frac{d\lambda_k}{dt}(t)\cdot
\partial_\lambda
S_{11}(t,\lambda_k(t))=0\,.
\end{equation}
Using the equation of motion for $\bf S$, one therefore finds
\begin{equation}
i\frac{d\lambda_k}{dt}(t)=\frac{1}{\partial_\lambda S_{11}(t,\lambda_k(t))}
\int_{-\infty}^\infty\left[{\bf M}^+(x',t,\lambda_k(t))^{-1}{\bf P}
{\bf M}^-(x',t,\lambda_k(t))\right]_{11}dx'\,.
\label{eq:eigenvalueevolve}
\end{equation}
The integrand here is 
$
p[\psi,\psi^*]M^+_{22}M^-_{21}-
p[\psi,\psi^*]^*M^+_{12}M^-_{11}
$,
evaluated at $x'$, $t$, and $\lambda=\lambda_k(t)$.  As remarked
above, this makes sense with $\Im(\lambda_k(t))>0$.
It remains to find an equation for $\{\gamma_k(t)\}$.
Differentiating the defining relation $M_2^+(x,t,\lambda_k(t))=
\gamma_k(t)M_1^-(x,t,\lambda_k(t))$ with respect to
$t$ and using the evolution equation for 
${\bf M}$ taken in the limit
$\lambda\rightarrow\lambda_k(t)$, yields the equation of motion
\begin{equation}
\begin{array}{rcl}\displaystyle
\left[i\frac{d\gamma_k}{dt}(t)-2\lambda^2\gamma_k(t)\right]
M_1^-(x,t,\lambda)&=&\displaystyle
\gamma_k(t)\int_{-\infty}^\infty {\bf H}_k(x,x',t)
{\bf P}M_1^-(x',t,\lambda)\,dx' + \\\\
&&\displaystyle
i\left[\partial_\lambda M_2^+(x,t,
\lambda)-\gamma_k(t)\partial_\lambda M^-_1(x,t,\lambda)\right]
\frac{d\lambda_k}{dt}(t)\,,\end{array}
\label{eq:gammaevolve}
\end{equation}
with $\lambda=\lambda_k(t)$.
The equations (\ref{eq:reflectionevolve}),
(\ref{eq:eigenvalueevolve}), and (\ref{eq:gammaevolve}) 
describe the evolution of the scattering data, but are
coupled to the equations for ${\bf
M}$ and ${\bf M}^\pm$.  This coupling disappears for
${\bf P}\equiv{\bf 0}$:
\begin{equation}
i\partial_t S_{12}(t,\lambda) - 2\lambda^2 S_{12}(t,\lambda)=0\,,
\hspace{0.3 in}
i\frac{d\lambda_k}{dt}(t)=0\,,\hspace{0.3 in}
i\frac{d\gamma_k}{dt}(t)-2\lambda_k(t)^2\gamma_k(t)=0\,,
\end{equation}
for $k=1,\dots,N$, as was first observed by Zakharov and Shabat
\cite{Zakharov:1972}.  From this simple system, it is possible to
introduce the coupling perturbatively, leading to closed systems order
by order.

\section{Perturbation Theory with Nearly Bound Solitons} 
\label{sec:perturb}
We now suppose that $p[\psi,\psi^*]=\epsilon^2W(|\psi|^2)\psi$ for
some real-valued function $W$, taking $\epsilon>0$ to be a small
parameter, and seek a perturbative solution of the equations of motion
for the scattering data.  We want a description of the solution up to
an $O(\epsilon^2)$ error, containing important physical information,
and valid uniformly over time scales of length
$O(\epsilon^{-1})$.  The initial data we consider is
\begin{equation}
S_{12}(0,\lambda)\equiv 0\,,\hspace{0.3 in}
\lambda_k(0)=im_k\,,\hspace{0.3 in}
\gamma_k(0)=\exp(-2m_kx_k^0+i\theta_k^0)\,.
\label{eq:data}
\end{equation}
\begin{proposition}
The solution of the initial-value problem of
(\ref{eq:reflectionevolve}), (\ref{eq:eigenvalueevolve}), and
(\ref{eq:gammaevolve}) with
initial conditions (\ref{eq:data}), is given asymptotically for small
$\epsilon$ by $S_{12}(t,\lambda)=O(\epsilon^2)$ and
\begin{equation}
\lambda_k(t)
=-\frac{\epsilon}{2}v_k(\epsilon t) + im_k + 
O(\epsilon^2)\,,
\hspace{0.3 in}
\gamma_k(t)
=\exp(-2m_kx_k(\epsilon t)+i\theta_k(t)+O(\epsilon^2))\,,
\end{equation}
where $x_k(T)$, $v_k(T)$, and $\theta_k(t)$ are certain functions to
be specified below.  They satisfy $x_k(0)=x_k^0$, $v_k(0)=0$ and
$\theta_k(0)= \theta_k^0$.  This approximation is uniformly
valid for times $t=O(\epsilon^{-1})$.
\end{proposition}

We develop the expansion using the multiscale formalism.  Introducing
the slow time variable $T=\epsilon t$, and assuming all quantities to
depend functionally on both $t$ and $T$, we replace the time derivatives
in (\ref{eq:reflectionevolve}), (\ref{eq:eigenvalueevolve}), and
(\ref{eq:gammaevolve}) according to the chain rule: $
\partial_t\rightarrow \partial_t + \epsilon\partial_T $.  Observe that
for the initial conditions (\ref{eq:data}), there is no enforced
magnitude for $\Re(\lambda_k)$ or $S_{12}(\lambda)$.  We may thus
select the scaling of these quantities to achieve a dominant balance.
We choose to scale $S_{12}(\lambda)$ as
$\epsilon^2$ and $\Re(\lambda_k)$ as $\epsilon$.  Thus, setting
$\lambda_k=\epsilon a_k + ib_k$ and $\gamma_k=\exp(\Delta_k+i\xi_k)$,
we assume the expansions:
\begin{equation}
\begin{array}{rcl}
\epsilon a_k + ib_k&=&
\epsilon (a_k^{(0)}+\epsilon a_k^{(1)} +\dots) + i(b_k^{(0)}+ \epsilon b_k^{(1)}+\dots )\,,\\\\
S_{12}&=&\epsilon^2 (S_{12}^{(0)} + \epsilon S_{12}^{(1)}+\dots)\,,\\\\
\Delta_k+i\xi_k&=&(\Delta_k^{(0)}+\epsilon\Delta_k^{(1)}+\dots) + 
i(\xi_k^{(0)} + \epsilon \xi_k^{(1)}+\dots)\,.
\end{array}
\end{equation}

Substituting into the equations of motion
and collecting
powers of $\epsilon$, we examine the resulting equations order by
order.  First, from the leading-order terms in
(\ref{eq:eigenvalueevolve}) we find for $k=1,\dots,N$ that
\begin{equation}
a_k^{(0)}=a_k^{(0)}(T)\,,\hspace{0.3 in}b_k^{(0)}=b_k^{(0)}(T)\,,
\end{equation}
so that these quantities do not depend on the fast time $t$.
Similarly, looking at (\ref{eq:gammaevolve}) we see that
\begin{equation}
\Delta_k^{(0)}=\Delta_k^{(0)}(T)\,,\hspace{0.3 in}
\xi_k^{(0)}=\xi_k^{(0)}(0) - 2b_k^{(0)}(T)^2t\,.
\label{eq:xiformula}
\end{equation}
The description we desire will follow upon determining the
$T$ dependence of these leading-order quantities.  The 
$O(\epsilon)$ contribution in the equation for $b_k$, the imaginary part
of (\ref{eq:eigenvalueevolve}), is
\begin{equation}
\partial_tb_k^{(1)}+\partial_Tb_k^{(0)}=0\,.
\label{eq:bkcorrection}
\end{equation}
If this equation for $b_k^{(1)}$ is to be solvable in the class of
bounded functions of $t$, then $b_k^{(0)}$ must be independent of $T$
as well as $t$.  With the $T$ dependence of $b_k^{(0)}$ dropped,
(\ref{eq:bkcorrection}) can be solved by taking $b_k^{(1)}=0$.  This
yields the simplest part of the claimed result, that $\Im( \lambda_k)$
is described uniformly for $t=O(\epsilon^{-1})$ by $ b_k(t)=m_k +
O(\epsilon^2)$, where the $m_k$ are constants.  Since $b_k^{(0)}=m_k$,
this also determines the leading-order behavior of $\xi_k^{(0)}$ from
(\ref{eq:xiformula}).  Setting $\theta_k^0:=\xi_k^{(0)}(0)$, we define
$\theta_k(t)$ as follows
\begin{equation}
\theta_k(t):= \xi_k^{(0)} = \theta_k^0 - 2m_k^2t\,.
\end{equation}
At 
$O(\epsilon)$, equation (\ref{eq:gammaevolve}) gives
\begin{equation}
\partial_t\Delta_k^{(1)} + \partial_T\Delta_k^{(0)} = 4a_k^{(0)}b_k^{(0)}=4m_ka_k^{(0)}\,.
\end{equation}
Again, avoid secular growth of $\Delta_k^{(1)}$ by setting
\begin{equation}
\partial_T\Delta_k^{(0)}(T)=4m_ka_k^{(0)}(T)\,,
\label{eq:Delta}
\end{equation}
and then take $\Delta_k^{(1)}=0$.  If we now define:
\begin{equation}
x_k(T):= -\frac{\Delta_k^{(0)}(T)}{2m_k}\,,\hspace{0.3 in}
v_k(T):= -2a_k^{(0)}(T)\,,
\end{equation}
then (\ref{eq:Delta}) takes the simple form:
\begin{equation}
x_k'(T)=v_k(T)\,.
\label{eq:x(T)}
\end{equation}

An equation for $v_k(T)$ is found at
$O(\epsilon^2)$ in the real part of
(\ref{eq:eigenvalueevolve}).  We find
\begin{equation}
\partial_t a_k^{(1)} -\frac{1}{2} v'(T) = \Im(G_k(t,T))\,,
\end{equation}
where $G_k(t,T)$ is the leading term, divided by $\epsilon^2$, of the
right hand side of (\ref{eq:eigenvalueevolve}).  In more detail, from
(\ref{eq:HilbertXF}) and the leading-order behavior of 
$\{\lambda_k\}$, we first see that
\begin{equation}
\partial_\lambda S_{11}(t,\lambda_k(t))\Bigg|_{\epsilon=0}=
\partial_\lambda\prod_{j=1}^N\frac{\lambda-im_j}{\lambda+im_j}\Bigg|_{\lambda=
im_k}=\frac{1}{2im_k}\prod_{j\neq k}\frac{m_k-m_j}{m_k+m_j}\,.
\end{equation}
To find the leading-order behavior of 
${\bf M}^\pm$, recall that
$S_{12}=O(\epsilon^2)$ so that 
we can use the ``reflectionless'' construction of ${\bf M}^\pm$
and $\psi$ in terms of ${\bf
U}$, which in turn is constructed from 
$\{\lambda_k\approx im_k\}$ and 
$\{\gamma_k\approx\exp(-2m_kx_k(T)+i(\xi_k^{(0)}(0)+2m_kt))\}$.  
This gives
\begin{equation}
G_k(t,T)=i(-1)^N\left[2m_k\prod_{j\neq k}(m_k^2-m_j^2)\right]^{-1}
\int_{-\infty}^\infty W(|\psi(x,t)|^2)f(x,t,im_k)\,dx\,,
\label{eq:Gformula}
\end{equation}
with
\begin{equation}
f(x,t,\lambda):= \psi(x,t)U_{22}(x,t,\lambda)
U_{21}(x,t,\lambda)-\psi(x,t)^*U_{12}(x,t,\lambda)U_{11}(x,t,\lambda)\,.
\label{eq:fdef}
\end{equation}
Now, it is clear from (\ref{eq:relations}) that all of the $x$ and $t$
dependence in ${\bf U}$ and $\psi$ enters through the products
$\gamma_k\exp(-2i\lambda_k x)\approx\exp(\zeta_k)\exp(i\theta_k(t))$,
where $ \zeta_k:= 2m_k(x-x_k(T))$.  Therefore, $G_k(t,T)$ is a
multiperiodic function of $t$ for fixed $T$.  The $N-1$ frequencies
are independent of $T$, since all of the $T$ dependence enters through
the functions $x_k(T)$.  Secular growth of $a_k^{(1)}(t)$ is avoided
by choosing $v_k'(T)$ to cancel the mean value of this oscillatory
function:
\begin{equation}
m_kv_k'(T)= F_k(x_1(T),\dots,x_N(T)):=-2m_k\left\langle \Im(G_k(\cdot,T))\right\rangle\,,
\label{eq:v(T)}
\end{equation}
where angled brackets denote averaging over $t$ with $T$ 
fixed.  The force functions $F_k$ depend parametrically on the masses
$m_k$.  Equations (\ref{eq:x(T)}) and (\ref{eq:v(T)}) imply Newton's
equations for a system of interacting particles of mass $m_k$ and
coordinate $x_k$:
\begin{equation}
m_kx_k''(T)=F_k(x_1(T),\dots,x_N(T))\,.
\label{eq:Newton}
\end{equation}
It is easy to see that
$F_k(x_1+dx,x_2+dx,\dots,x_N+dx)=F_k(x_1,x_2,\dots,x_N)$ so that the
forces only depend on the relative coordinates.  There is also
a symmetry for (\ref{eq:Newton}) coming from the
conservation of momentum that holds exactly (and thus to all
orders of expansion) in 
(\ref{eq:fundamentalIVP}) with $p[\psi,\psi^*]=\epsilon^2
W(|\psi|^2)\psi$.  This symmetry follows from the trace formula
(\ref{eq:trace}) and shows that the total force on the system is
zero:
\begin{equation}
\sum_{k=1}^NF_k(x_1(T),\dots,x_N(T))=\sum_{k=1}^N m_kx_k''(T)=0\,.
\end{equation}

The dynamical system (\ref{eq:Newton}) describes the evolution of the
scattering data.  Since the reflection coefficient vanishes to second
order on the time scales of interest, solutions of
(\ref{eq:Newton}) can be used to build, at each fixed $t$, the
$N$-soliton potential as in \S\ref{sec:reflectionless}.  This allows a
direct comparison between numerics for (\ref{eq:fundamentalIVP}) and
the predictions of (\ref{eq:Newton}).

\section{Two Particles}
\label{sec:two}
Consider the case $N=2$.  The
aforementioned symmetries imply that the system takes the form
\begin{equation}
\begin{array}{rcccl}
m_1x_1''(T)&=&F_1(x_1(T),x_2(T))&=&-\frac{1}{2}F(x_2(T)-x_1(T))\,,\\
m_2x_2''(T)&=&F_2(x_1(T),x_2(T))&=&\frac{1}{2}F(x_2(T)-x_1(T))\,,
\end{array}
\end{equation}
for some function $F$.  The relevant quantity is then the relative
distance $y(T):=x_2(T)-x_1(T)$, which has the simple-looking equation
of motion
\begin{equation}
\tilde{m}y''=F(y)\,,
\end{equation}
where the {\em effective mass} is defined by 
$
\tilde{m}:= 2 \left(m_2^{-1}+m_1^{-1}\right)^{-1}$.

\subsection{Writing down the force function.}
We begin our study of the force functions by simplifying
the integrand in (\ref{eq:Gformula}) to isolate terms that are exact
$x$-derivatives and do not contribute.  In
this context, consider the {\em squared
eigenfunction system} implied by (\ref{eq:linsys}).
Let ${\bf M}$ be {\em any} solution of $\partial_x{\bf M}={\bf
LM}$, and define the quadratic forms
\begin{equation}
\phi:=  M_{11}M_{12}\,,\hspace{0.3 in}
\chi:=  M_{21}M_{22}\,,\hspace{0.3 in}
\eta:=  M_{11}M_{22}+
M_{12}M_{21}\,.
\end{equation}
Then, these quantities again satisfy a linear system of equations
\begin{equation}
\partial_x\phi=-2i\lambda\phi + 
\psi\eta\,,\hspace{0.3 in}
\partial_x\chi=2i\lambda\chi-\psi^*
\eta\,,\hspace{0.3 in}
\partial_x\eta=-2\psi^*\phi+
2\psi\chi\,.
\end{equation}
Using the quadratic forms associated with  
${\bf U}$, $f$ as defined
by (\ref{eq:fdef}) is seen to be an exact $x$-derivative:
\begin{equation}
f= \frac{1}{2}\partial_x\eta
=
\frac{1}{2}\partial_x(U_{11}U_{22}+
U_{12}U_{21})
=\partial_x(U_{12}U_{21})\,,
\label{eq:fsimp}
\end{equation}
where the last equality follows from the fact that the determinant of
any solution of (\ref{eq:linsys}) is independent of $x$ because
${\bf L}$ is traceless.  For $N=2$, 
we use the relations (\ref{eq:relations}) 
and the parameters $\lambda_1=im_1$, $\lambda_2=im_2$, 
$\gamma_1=\exp(-2m_1x_1(T)+i(\theta_1^0-2m_1^2t))$, and
$\gamma_2=\exp(-2m_2x_2(T)+i(\theta_2^0-2m_2^2t))$ to find
$
U_{12}=e^{i\lambda x}(\lambda \psi/(2i) +
\varphi)$ and 
$U_{21}=e^{-i\lambda x}(\lambda\psi^*/(2i) -
\varphi^*)$,
where
\begin{equation}
\begin{array}{rcl}
\psi&=&\displaystyle
\frac{2(m_2^2-m_1^2)}{D(\zeta_1,\zeta_2,\theta_2-\theta_1))}\left[m_1\cosh(\zeta_2)e^{i\theta_1(t)}-m_2\cosh(\zeta_1)e^{i\theta_2(t)}\right]\,,\\\\
\varphi&=&\displaystyle
\frac{m_1m_2(m_2^2-m_1^2)}{D(\zeta_1,\zeta_2,\theta_2-\theta_1)}\left[
\sinh(\zeta_2)e^{i\theta_1(t)}-\sinh(\zeta_1)e^{i\theta_2(t)}\right]\,,
\end{array}
\label{eq:quantities}
\end{equation}
where $\psi$ is the well-known two-soliton ``breather'' solution, and using
\begin{equation}
D(\zeta_1,\zeta_2,\theta):=(m_1+m_2)^2\cosh(\zeta_1)\cosh(\zeta_2)-
2m_1m_2
\cosh(\zeta_1+\zeta_2)
-2m_1m_2\cos(\theta)\,.
\end{equation}

Since $W(\cdot)\in{\mathbb R}$, only $\Re(f(x,t,im_k))$ 
is needed to find $\Im(G_k(t,T))$.  From (\ref{eq:fsimp}) one finds
$
f=-\partial_x(
\lambda^2|\psi|^2/4 + \lambda\Im(\psi\varphi^*) + |\varphi|^2)$,
and therefore
$
\Re(f(x,t,im_k))=m_k^2\partial_x|\psi|^2/4-\partial_x
|\varphi|^2
$.
Using this in the formula (\ref{eq:Gformula}) for $\Im(G_k(t,T))$, one
finds that the first term is an exact derivative of a rapidly
decreasing function and hence integrates away.  In terms of the two
quantities $|\psi|^2$ and $|\varphi|^2$ obtained directly from
(\ref{eq:quantities})
we finally obtain
\begin{equation}
\Im(G_1(t,T))=
\frac{1}{2m_1(m_2^2-m_1^2)}\int_{-\infty}^\infty
W(|\psi|^2)\partial_x|\varphi|^2\,dx
=-\frac{m_2}{m_1}
\Im(G_2(t,T))\,.
\end{equation}
In particular, it follows that
$-2m_1\Im(G_1(t,T))-2m_2\Im(G_2(t,T))=0$ so that the total
instantaneous (that is, before averaging over $t$) force vanishes.

Specializing further to the quintic perturbation
(\ref{eq:quintic}) by taking $ W(\rho):=\sigma\rho^2$ and writing
\begin{equation}
F(y;m_1,m_2)=\frac{1}{2\pi}\int_0^{2\pi}\int_{-\infty}^\infty h(y,z,\theta;m_1,m_2)\,dz\,d\theta\,,
\label{eq:Fexpression}
\end{equation}
we have found the following explicit formula for $h$:
\begin{equation}
h(y,z,\theta;m_1,m_2)=
\frac{128\sigma m_1^2m_2^2(m_2^2-m_1^2)^5}{D(\zeta_1,\zeta_2,\theta)^7}\left[
h_1+\dots+h_{13}\right]\,,
\label{eq:eqnforh}
\end{equation}
where the individual terms $h_k$ are given in the Appendix.  They
depend on a dummy integration variable $z$ that differs from $x$ by a
simple translation.  Note that, by the periodicity 
with respect to the ``fast'' function
$\theta=\theta_2(t)-\theta_1(t)$, averaging over $t$ is equivalent to
averaging over $\theta$.

\subsection{Scale invariance.}
From (\ref{eq:Fexpression}) and the explicit formulae for
the terms $h_k$ in the Appendix, note the
important symmetry:
\begin{equation}
F(\xi y; m_1, m_2)=\xi^{-6} F(y;\xi m_1,\xi m_2)\,,
\end{equation}
for all nonzero $\xi\in{\mathbb R}$.  Setting $y=\xi q$ and
$S=\xi^{-3} T$, the equation of motion takes the form:
\begin{equation}
(\xi\tilde{m})q''(S)=F(q(S);\xi m_1,\xi m_2)\,.
\end{equation}
For arbitrary masses $m_1$ and $m_2$, we may then set
$\xi=(m_1m_2)^{-1/2}$.  Because $\tilde{m}$ is homogeneous
of degree one in $m_1$ and $m_2$, it is convenient to use the
normalized masses 
\begin{equation}
M_1=\xi m_1\,,\hspace{0.3 in}M_2=\xi m_2\,,\hspace{0.3 in}
\tilde{M}=\xi\tilde{m}\,,\hspace{0.3 in}\xi=(m_1m_2)^{-1/2}\,.
\end{equation}
Here, $M_1$ and $M_2$ satisfy $M_1M_2=1$ and may therefore be expressed
in terms of the 
{\em normalized effective mass} $\tilde{M}$ by solving
$2(M_2^{-1}+M_1^{-1})^{-1}=\tilde{M}$
subject to this constraint to find:
\begin{equation}
M_1=\left[1-(1-\tilde{M}^2)^{1/2}\right]\cdot\tilde{M}^{-1}\,,
\hspace{0.3 in}
M_2=\left[1+(1-\tilde{M}^2)^{1/2}\right]\cdot\tilde{M}^{-1}\,,
\end{equation}
assuming without loss of generality that $M_2>M_1$.  From
now on, we will work exclusively with the normalized masses, in which
case the force depends only on $S$ and $\tilde{M}$.

\subsection{Averaging.}
We now compute the $\theta$-averages 
explicitly by residues.  There are five terms:
\begin{equation}
A_p:= 
\frac{1}{2\pi}
\int_0^{2\pi}\frac{\cos^p\theta}{D(\zeta_1,\zeta_2,\theta)^7}\,d\theta
=
\frac{2^{-7}}{2\pi}
\int_0^{2\pi}\frac{\cos^p\theta}{(a-\cos\theta)^7}\,d\theta\,,
\end{equation}
for $p=0,1,\dots,4$, where $ a:=
(2-\tilde{M}^2)\cdot\cosh(\zeta_1)\cosh(\zeta_2)\cdot\tilde{M}^{-2}-\sinh(\zeta_1)\sinh(\zeta_2)
\ge 1$.  Changing variables to
$w=\exp(i\theta)$, the contour of integration becomes the
clockwise-oriented unit circle in the $w$-plane.  The only
singularity within the contour is a seventh-order
pole at the point $w_0= a-(a^2-1)^{1/2}$, where from here on the
positive root is taken.  Therefore,
\begin{equation}
A_p=-\frac{1}{2^p}\begin{array}{c}{\rm Res}\\{\scriptstyle w=w_0}
\end{array}\frac{w^6(w+w^{-1})^p}
{(w-w_0)^7(w-w_0^{-1})^7}\,.
\end{equation}
In particular one finds exact expressions for $\tilde{A}_p\doteq
65536(a^2-1)^{13/2}A_p$:,
\begin{equation}
\begin{array}{rclrcl}
\tilde{A}_0&=&8(2a)^6+240(2a)^4+720(2a)^2+160
\,,&
\tilde{A}_1&=&56(2a)^5+560(2a)^3+560(2a)
\,,\\
\tilde{A}_2&=&4(2a)^6+232(2a)^4+808(2a)^2+192\,,
&\tilde{A}_3&=&42(2a)^5+588(2a)^3+672(2a)\,,\\
\tilde{A}_4&=&3(2a)^6+202(2a)^4+928(2a)^2+256\,.
\end{array}
\end{equation}
These results yield an explicit formula for the
two-particle force function as an integral
\begin{equation}
F(q;\tilde{M})=\int_{-\infty}^\infty H(q,z)\,dz\,,
\label{eq:forceformula}
\end{equation}
where we are assuming that $M_1M_2=1$ and $M_2>1>M_1>0$, and where
\begin{equation}
\begin{array}{rcl}
H&=&\displaystyle
\frac{2\sigma (1-\tilde{M}^2)^{5/2}}{\tilde{M}^{10}}\sum_{m,n=0}^6
\left[g_{mn}(M_1,M_2)\tanh\zeta_1 + g_{nm}(M_2,M_1)\tanh\zeta_2\right]H_{mn}\,,\\\\
H_{mn}&:=&\displaystyle
\frac{{\,\rm sech\,}^{2(6-m)}\zeta_1{\,\rm sech\,}^{2(6-n)}\zeta_2}
{\left(\displaystyle \left(\frac{2-\tilde{M}^2}{\tilde{M}^2}-\tanh\zeta_1\tanh\zeta_2\right)^2-
{\,\rm sech\,}^2\zeta_1{\,\rm sech\,}^2\zeta_2\right)^{13/2}}\,.
\end{array}
\label{eq:eqnforH}
\end{equation}
Here, $\zeta_1=2M_1(z+q/2)$ and $\zeta_2=2M_2(z-q/2)$.  Many of the
coefficients $g_{mn}(\alpha,\beta)$ vanish identically.  In
particular, $g_{66}=0$ as is needed for the integral to converge. The
nonvanishing coefficients $g_{mn}(\alpha,\beta)$ are given in the
Appendix.

\subsection{General features of the force function.}
Unfortunately, (\ref{eq:forceformula}) cannot be evaluated in
closed form because the integrand generally involves both
$\exp(\zeta_1)$ and $\exp(\zeta_2)$.  Even if
$M_2/M_1\in{\mathbb Q}$ so that the integrand becomes
a rational function of, say, $\exp(\zeta_1)$,
the denominator 
is irreducible already for the simplest resonance, $M_2=2M_1$.

\begin{figure}[h]
\begin{center}
\mbox{\psfig{file=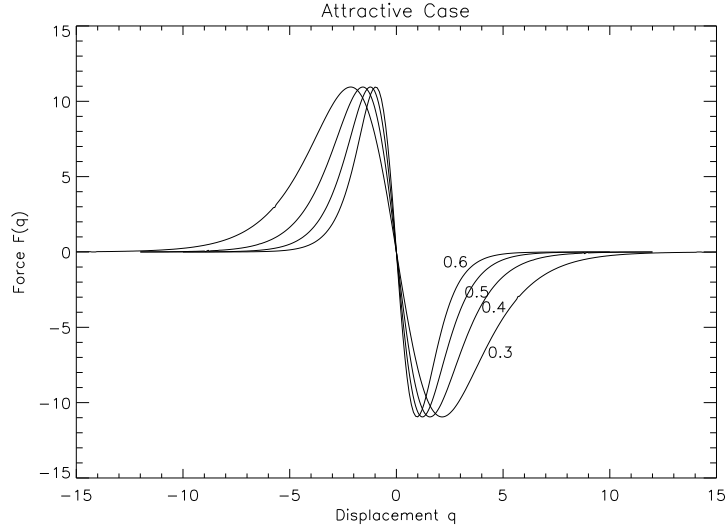,width=4 in}}
\end{center}
\caption{\em The force law $F(q,\tilde{M})$ in the attractive case,
$\sigma=+1$, for $\tilde{M}=0.3, 0.4, 0.5, 0.6$.  In the repulsive
case $\sigma=-1$, the force simply has the opposite sign.}
\label{fig:attractiveforce}
\end{figure}

In spite of these difficulties, certain elementary features of the
force law can be extracted:
\begin{itemize}
\item
$F(q;\tilde{M})$ is proportional to the constant $\sigma=\pm 1$, as is
clear from 
(\ref{eq:forceformula}).
\item
$F(q;\tilde{M})$ is an odd function of $q$, since
the integrand satisfies $H(-q,z)=-H(q,-z)$ and 
moreover this symmetry holds term by term in the formula for $H$.  
\item
$F(q;\tilde{M})$ decays to zero for large $q$.  This follows from the
fact that the denominator of each term $H_{mn}$ in the integral is
bounded and the corresponding numerator vanishes for large $q$ whenever
$g_{mn}\neq 0$.  The result then follows from 
a dominated-convergence argument.
\item
$F(q;\tilde{M})$ only vanishes exactly for $q=0$.  Thus it is strictly
of one sign for $q>0$.
\item
The normalized effective mass $\tilde{M}$ enters the dynamics both as
a mass parameter multiplying the acceleration $q''(S)$ and as a
parameter in $F(q;\tilde{M})$ itself.
\end{itemize}
The force $F(q;\tilde{M})$, as computed from the integral formula
(\ref{eq:forceformula}), is plotted in
Figure~\ref{fig:attractiveforce} for several different values of the
normalized effective mass $\tilde{M}$.

\subsection{Attractive case.  Spring constant.}
\begin{figure}[h]
\begin{center}
\mbox{\psfig{file=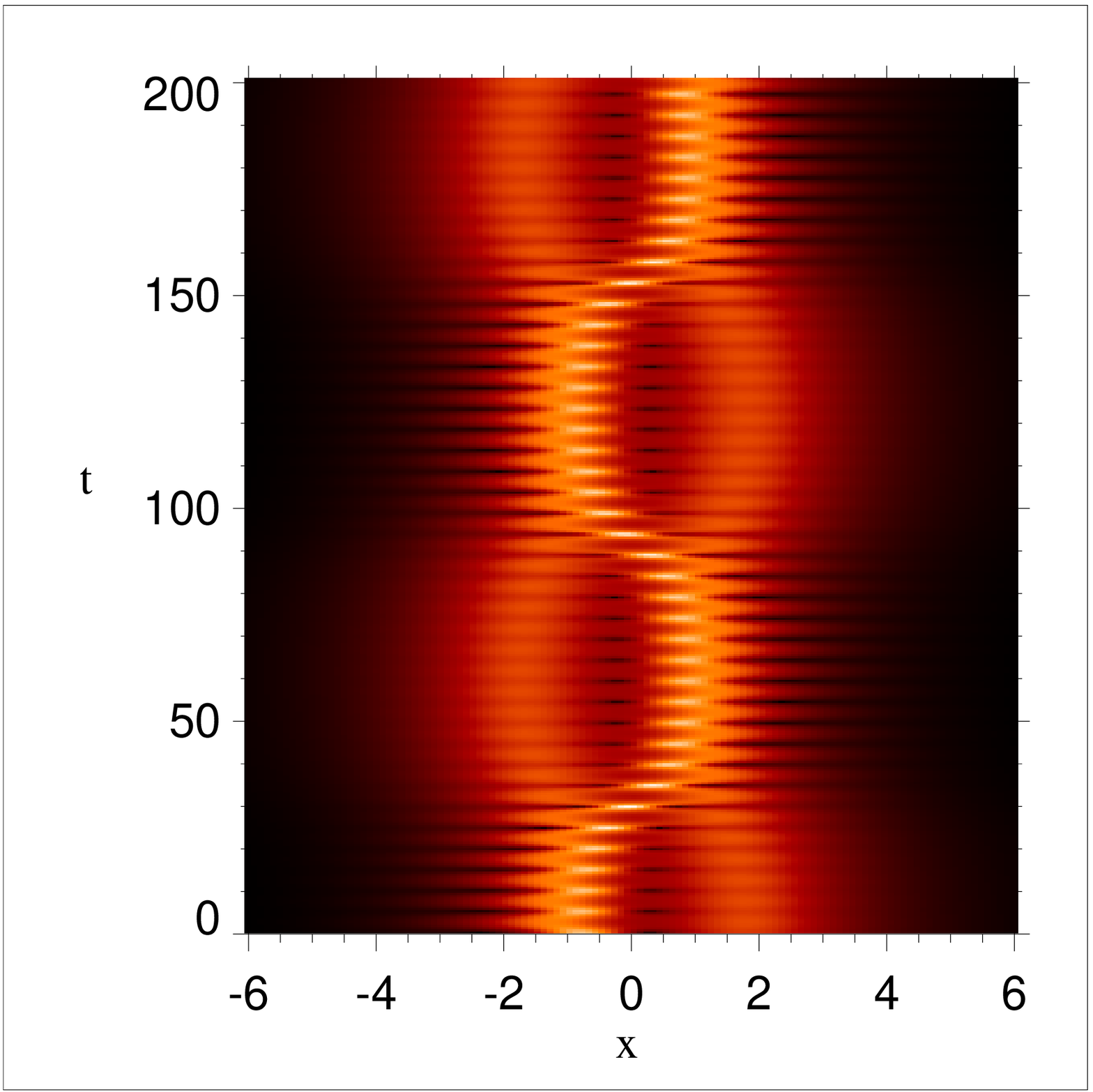,width=2.5 in}}
\mbox{\psfig{file=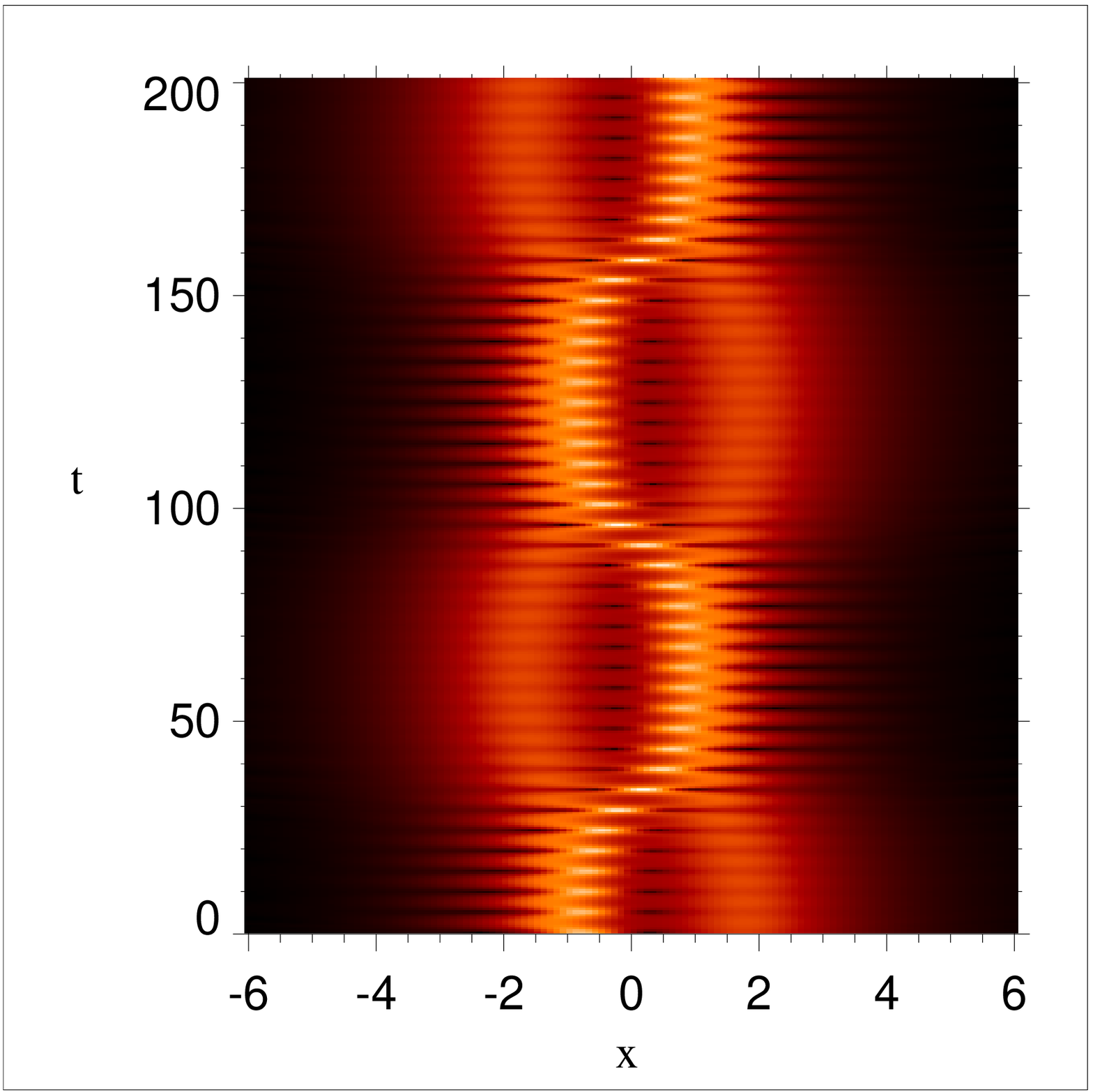,width=2.5 in}}
\end{center}
\caption{\em A two-soliton bound state affected by an attractive
perturbation.  Here, $\epsilon=0.0387$, $m_1=0.6$, and
$m_2=1$.  Left: an approximation to $|\psi|^2$ found by solving
Newton's equations
and then constructing the field using reflectionless inverse theory.
Right: the corresponding numerics for (\ref{eq:fundamentalIVP}).
The bound state has too much energy for the harmonic approximation to
hold, and the period of motion, about $120$ time units, is
longer than the harmonic
period.  }
\label{fig:bound}
\end{figure}

\begin{figure}[h]
\begin{center}\mbox{\psfig{file=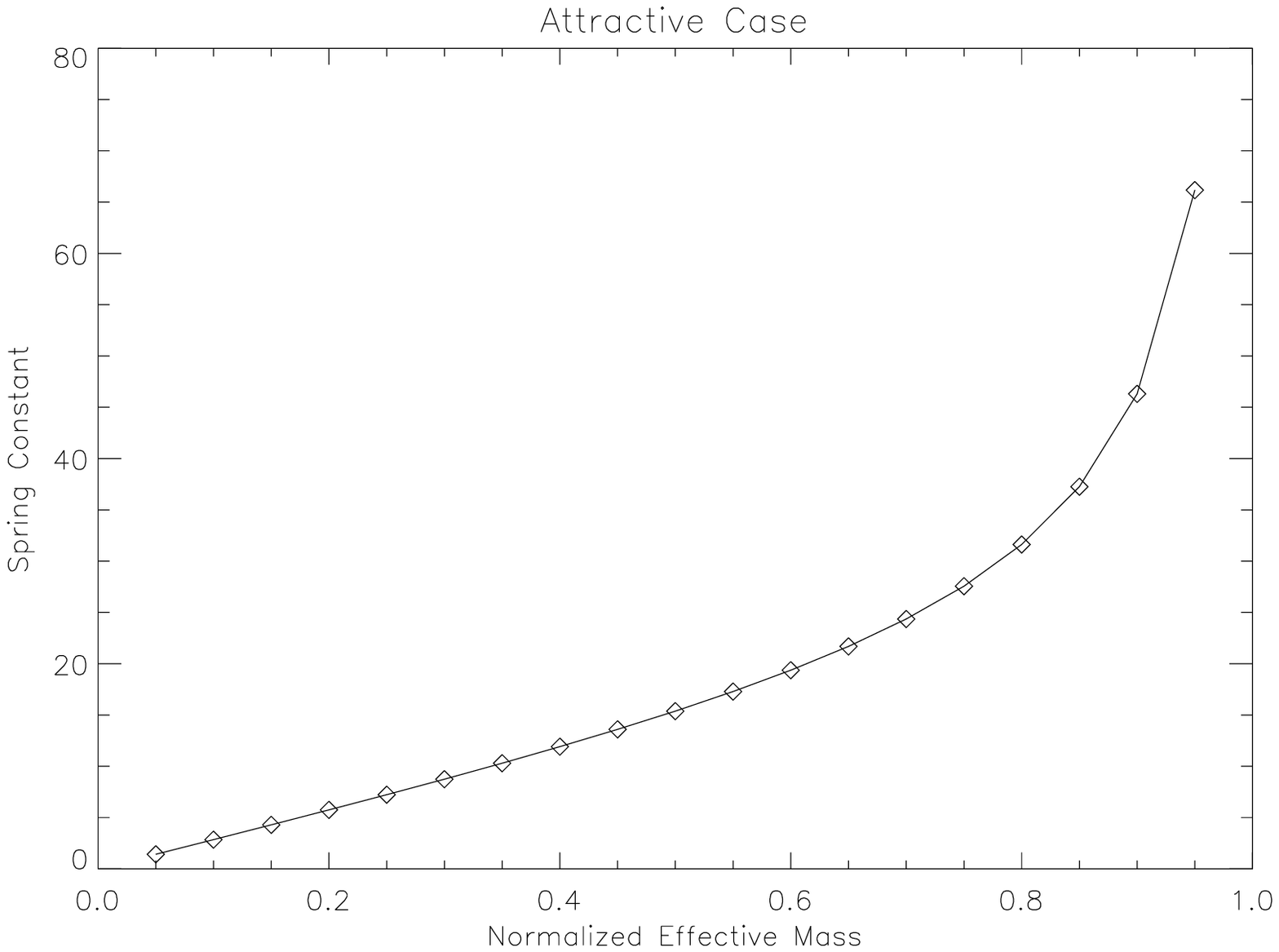,width=4 in}}\end{center}
\caption{\em The spring constant for small bound motions
as a function $\tilde{M}$.}
\label{fig:springconstant}
\end{figure}

\begin{figure}[h]
\begin{center}\mbox{\psfig{file=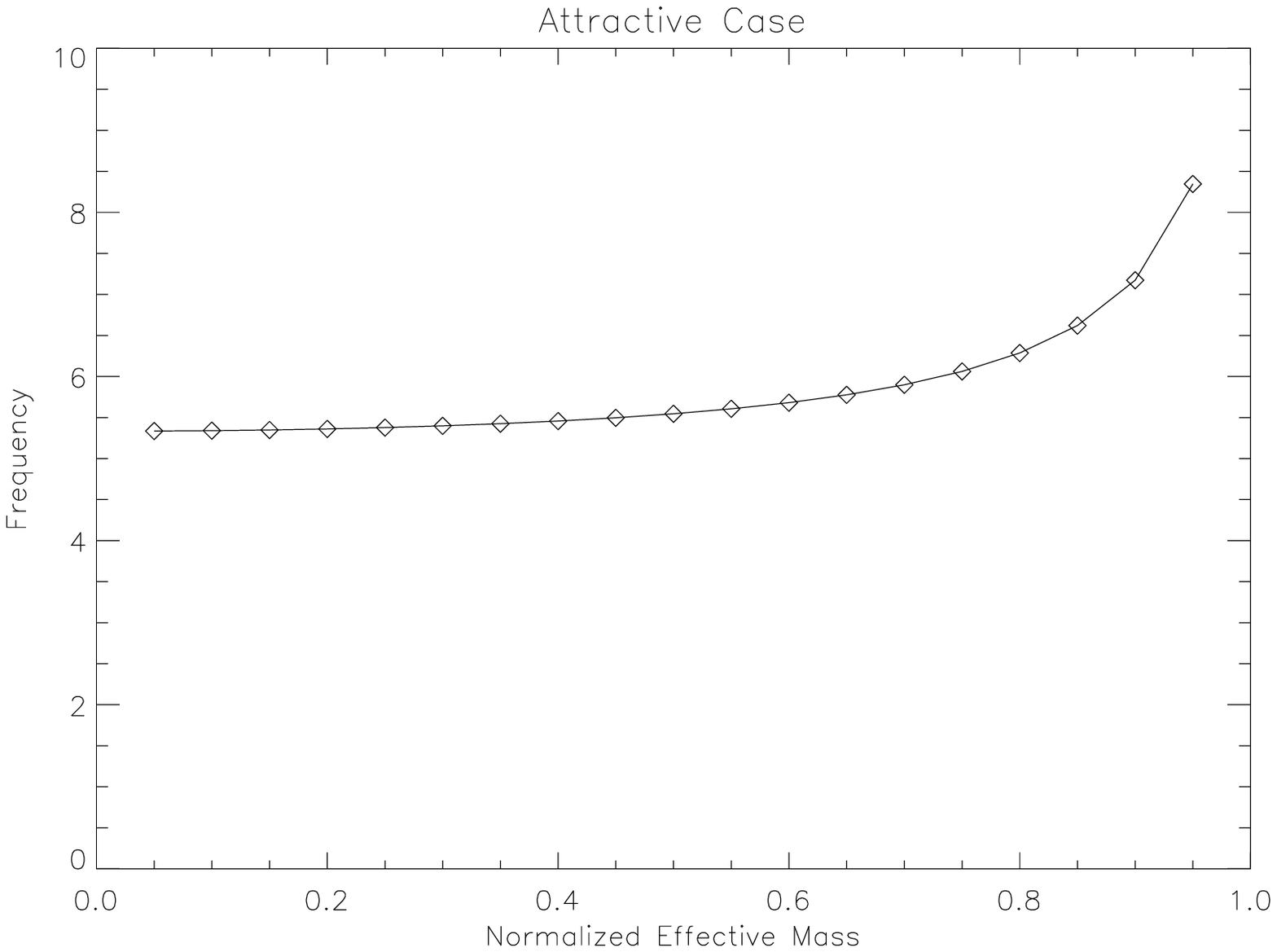,width=4 in}}\end{center}
\caption{\em The frequency $\omega$ of harmonic motion as a
function $\tilde{M}$.  
}
\label{fig:frequency}
\end{figure}

For $\sigma=+1$, the force $F(q;\tilde{M})$ and the displacement $q$
have opposite signs, so the force is always attractive.  This means
that the slow dynamics of the two-soliton bound state are periodic in
time and the state remains bound\footnote{This is a long-time
statement, holding for $t=O(\epsilon^{-1})$, but not an infinite time
statement.  The question of whether {\em true} breather-like bound
states exist (that is, permanently) for nonzero $\epsilon$ is more
subtle.}.  To illustrate,
Figure~\ref{fig:bound} compares the results of perturbation theory
with a simulation of (\ref{eq:fundamentalIVP}).
For small displacements, we have $F(q;\tilde{M})=-k(\tilde{M})q +
O(q^2)$.  The (mass-dependent) spring constant $k(\tilde{M})$
determines the frequency
$\omega(\tilde{M}):=(k(\tilde{M})/\tilde{M})^{1/2}$ of small
oscillatory motions.  This is the frequency on the time scale $S$;
the frequency on the original time scale $t$ is related by $
\Omega(m_1,m_2,\epsilon)=
\epsilon\,(m_1m_2)^{3/2}\omega(\tilde{m}/(m_1m_2)^{1/2})$.  A formula
for the spring constant $k(\tilde{M})$ can be found by simply
differentiating with respect to $q$ in
(\ref{eq:forceformula}) and setting $q=0$, however it seems less useful to
present than a plot, shown in
Figure~\ref{fig:springconstant}, of the
(numerically) evaluated formula.
In Figure~\ref{fig:frequency}
we plot the corresponding frequency (on the time scale $S$), the
latter being a directly observable quantity.
It is noteworthy here that the dynamics of solitons can be described
by a linear theory even though their amplitudes are not at all small.
The parameter linearizing the theory is the distance between the
solitons, rather than the soliton amplitude.  We also remark that the
limit in which this linear behavior holds is that of
infinitessimally-separated solitons, a limit in which methods assuming
the solitons to be well-separated are invalid.

\subsection{Repulsive case.  Asymptotic velocity.}
\begin{figure}[h]
\begin{center}
\mbox{\psfig{file=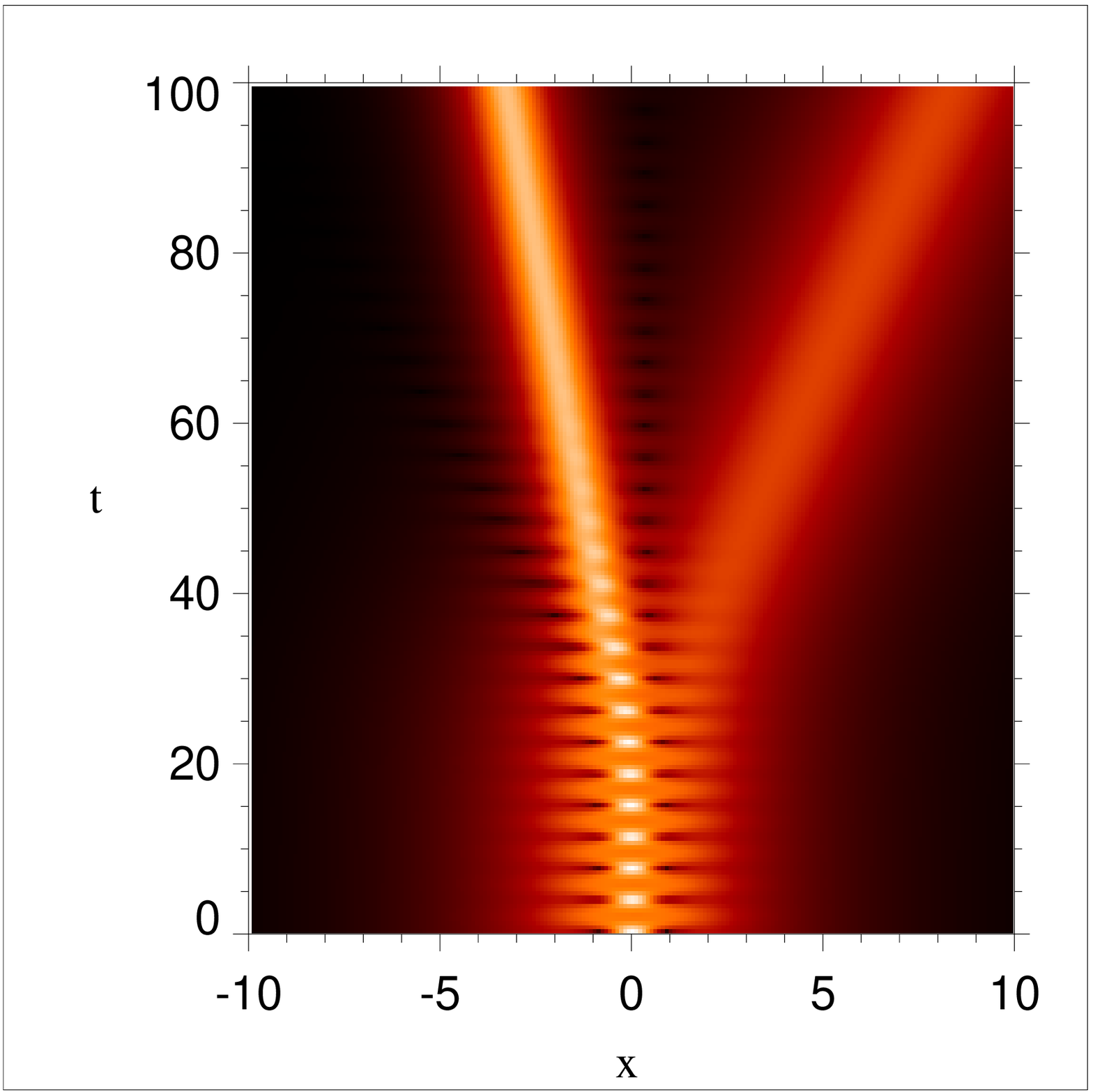,width=2.5 in}}
\mbox{\psfig{file=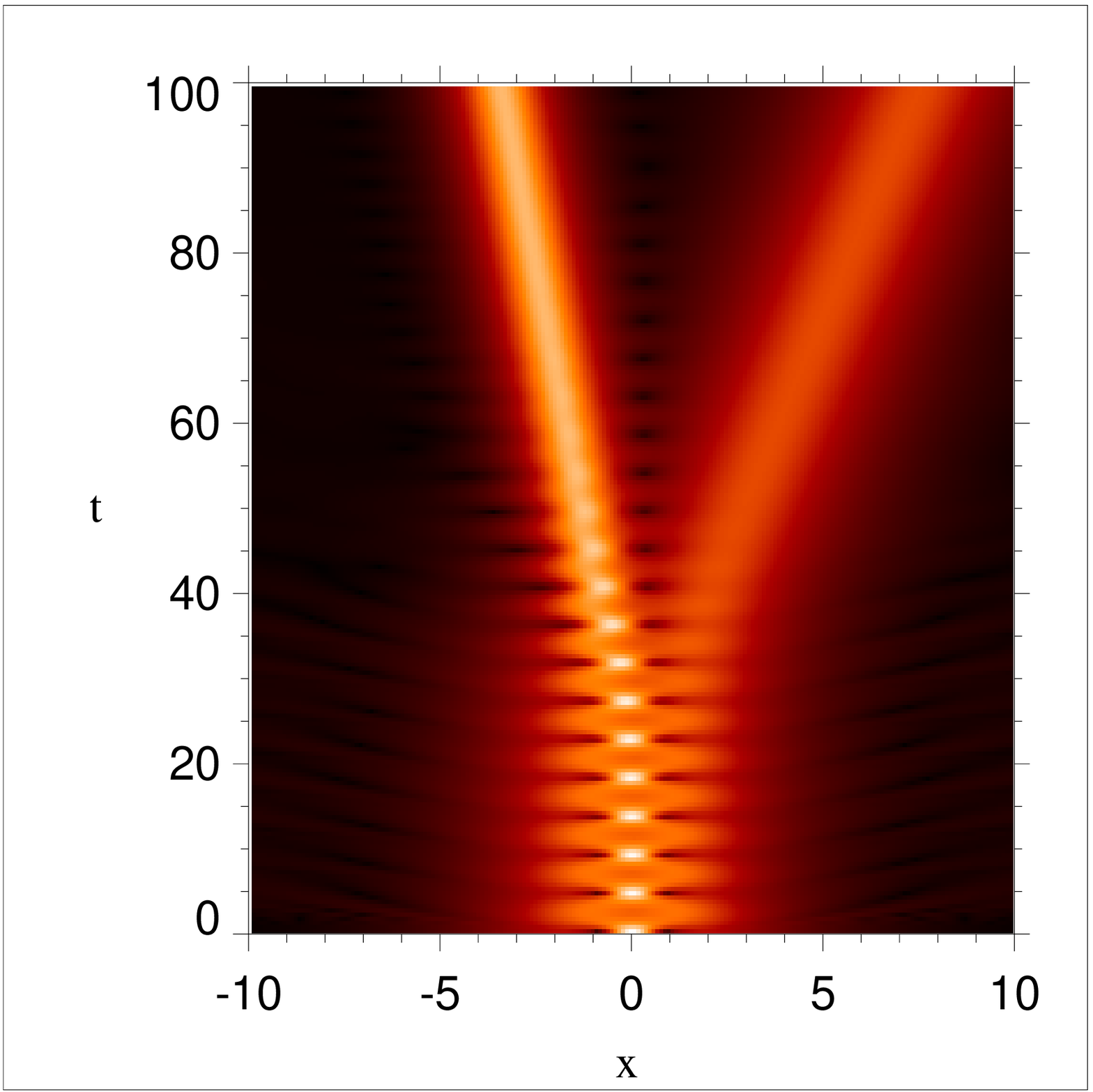,width=2.5 in}}
\end{center}
\caption{\em A two-soliton bound state affected by a repulsive
perturbation.  Here, $\epsilon=0.07746$, $m_1=0.4$, and $m_2=1$.  As
in Figure~\ref{fig:bound}, the result of perturbation theory is on the
left and the numerics
are on the right.  The solitons escape with a
relative ``ejection'' velocity given by
(\ref{eq:velocitydiff}).  }
\label{fig:splitter}
\end{figure}

\begin{figure}[h]
\begin{center}
\mbox{\psfig{file=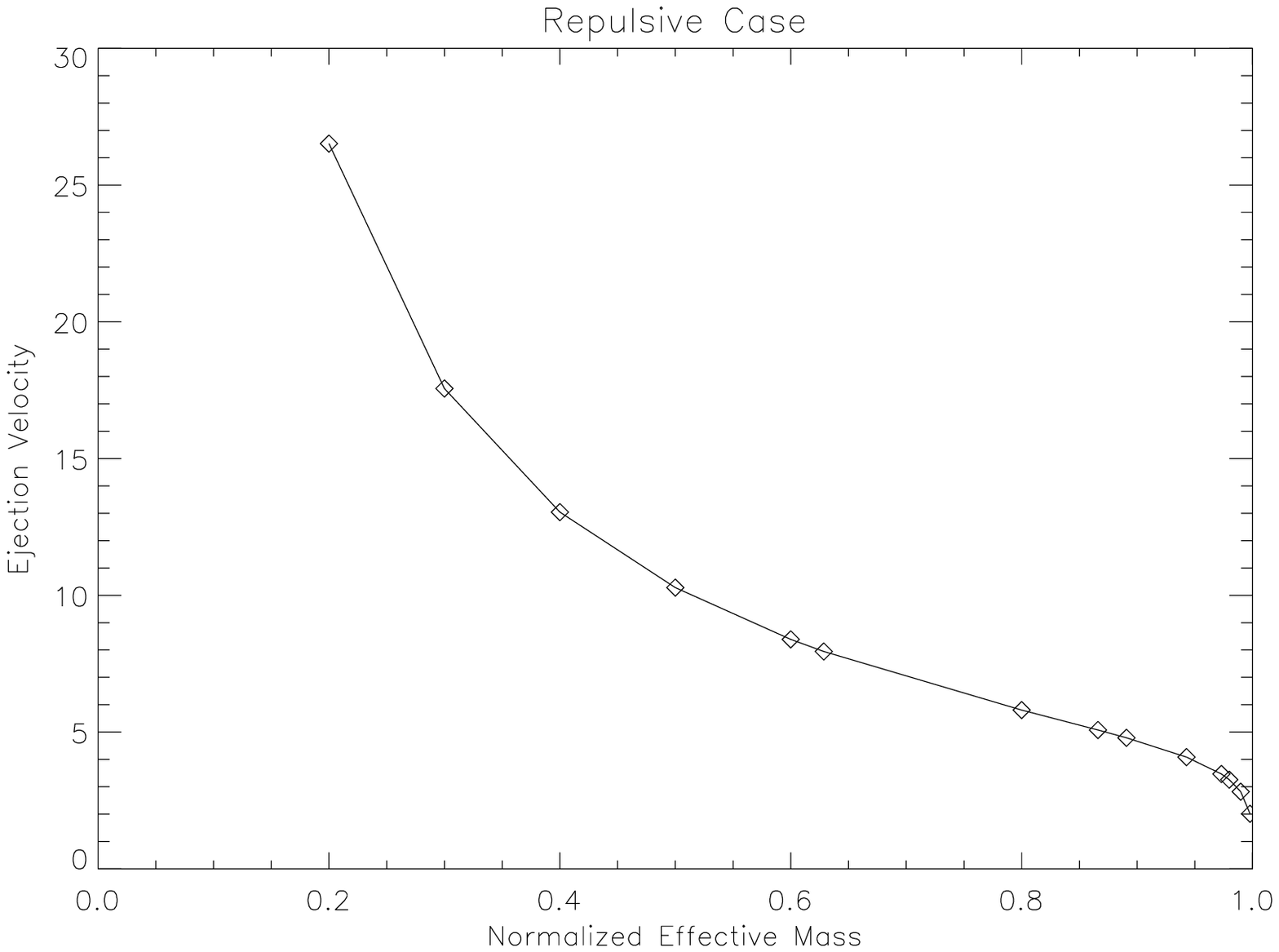,width=4 in}}
\end{center}
\caption{\em The asymptotic velocity difference $q'(\infty)$ of two
solitons falling from unstable equilibrium.
}
\label{fig:velocity}
\end{figure}

For $\sigma=-1$, the force and displacement $q$ have the same sign,
resulting in $q$ always becoming large.  Solitons that are near each
other at $t=0$ are ejected from the origin as observed by Artigas
et. al. \cite{Artigas:1997}.  This effect is captured accurately by
our theory, as shown in
Figure~\ref{fig:splitter}.
The work done by the force in moving the particle from $q=q_0$ to
$q=\infty$ determines the asymptotic velocity of an initially
stationary particle upon ejection.  Taking $q_0=0$ corresponds
to the ultimate velocity of a stationary particle that is slightly
perturbed from (unstable) equilibrium at the origin.  With zero
initial velocity, one equates the asymptotic kinetic energy with the work
done:
\begin{equation}
\frac{1}{2}\tilde{M}q'(\infty)^2=
\int_{q_0}^\infty
F(q;\tilde{M})\,dq\,,
\label{eq:binding}
\end{equation}
to find a formula for the asymptotic velocity difference:
\begin{equation}
q'(\infty)=\left(\frac{2}{\tilde{M}}\int_{q_0}^\infty \int_{-\infty}^\infty
H(q,z)\,dz\,dq\right)^{1/2}\,.
\label{eq:velocitydiff}
\end{equation}
Figure~\ref{fig:velocity} shows the asymptotic velocity difference
$q'(\infty)$ for $q_0=0$ found from (\ref{eq:velocitydiff}) as a
function of the normalized effective mass $\tilde{M}$.
To apply the graph in Figure~\ref{fig:velocity} to problems with
unnormalized masses, it is useful to unravel the changes
of variables made so far.  Given $m_1$ and $m_2$, the
scaling parameter is $\xi=(m_1m_2)^{-1/2}$ and the effective mass
is $\tilde{m}=2\cdot(m_1^{-1}+m_2^{-1})^{-1}$.  Then, the normalized
effective mass used in Figure~\ref{fig:velocity} is
$\tilde{M}=\xi\tilde{m}$.  Next, from the graph one finds the
asymptotic velocity $q'(\infty)$.  The true velocity in the original
coordinates is then $dy/dt=\epsilon\xi^{-2} q'(\infty)$.  For
example, the parameters used in Figure~\ref{fig:splitter} imply a
normalized effective mass of $\tilde{M}\approx 0.9$.  From 
Figure~\ref{fig:velocity} one finds $q'(\infty)\approx
5.0$, and thus $dy/dt\approx 0.15$.  This value agrees well with the
pictures in Figure~\ref{fig:splitter}.

In the attractive case, the integral (\ref{eq:binding})
also has physical meaning as the binding energy of the
two-soliton state.  A relative velocity in excess of
$q'(\infty)$, the escape velocity, will ``ionize'' the state.

\section{Discussion}
Multiscale asymptotics shows that under
certain conditions the behavior of a multisoliton initial condition
in a perturbed NLSE reduces to Newton's equations for a system of
interacting particles, one particle per soliton.  The theory
applies over time scales of length $O(\epsilon^{-1})$ for
perturbations of size $\epsilon^2$, when the initial velocities of the
solitons mutually differ by an $O(\epsilon)$ amount.  Our
calculations make very concrete the often-cited analogy between
solitons and particles.  We want to emphasize that the limit
considered here is one in which the relative velocities of the solitons
are small but the solitons may be strongly nonlinearly
superimposed, precisely the limit in which methods
exploiting large distances between solitons fail.

For a quintic perturbation of the NLSE and an initial condition
composed of two solitons, the resulting dynamical system can be
analyzed.  When the perturbation is attractive ($\sigma=+1$), the
system describes a nonlinear oscillator with all solutions $q(S)$
being periodic.  If the energy associated with $q(S)$ is small (that
is, if $q(0)$ and $q'(0)$ are both small), then the periodic motion is
nearly harmonic, and formulae for the associated spring constant and
frequency of motion can be found; in this limit the model for the
soliton interaction linearizes even though the soliton amplitudes are
not at all small.  The latter are determined by the masses $m_1$ and
$m_2$ and are not related to the coordinate $q(S)$.  For larger
energies, the spring ``softens'' and the frequency decreases
with increasing energy.  The pictures in Figure~\ref{fig:bound} show
oscillations in the nonlinear regime, where the frequency of motion is
smaller than the linear frequency.  Of course even in the nonlinear
regime, the dynamics still obey the simple model
$\tilde{M}q''=F(q;\tilde{M})$.  Although the periodic motion is
predicted and observed over long time scales of size
$O(\epsilon^{-1})$, it is not likely to persist for all time, due to
the influence of higher-order resonant coupling effects.

On the other hand, when the quintic perturbation is repulsive
($\sigma=-1$), the nodal point at the origin in the phase plane gets
replaced with an unstable saddle point.  All orbits apart from the
fixed point itself represent the nonlinear development of the
instability.  Because the force vanishes fast enough for large $q$,
the velocity $q'(S)$ ultimately saturates as the two-soliton state
becomes ``ionized''.  From the force function $F(q;\tilde{M})$ this
``ejection'' velocity may be calculated, giving excellent agreement
with direct simulations of the perturbed NLSE.  This analysis explains
the observations reported in \cite{Artigas:1997}.  The
symmetry-breaking that determines which soliton ends up
on the right and which on the left can be traced to
the location of the initial phase point in relation to the
separatrix connected to the saddle.  Unlike in the attractive
($\sigma=+1$) case, the approximation obtained from multiscale
asymptotics for the repulsive ($\sigma=-1$) case {\em is} expected to
be uniformly valid for all time, since as the solitons separate,
further effects due to resonant coupling diminish.

Given the formula for the force $F(q;\tilde{M})$, it is possible to
compute the harmonic frequency and ejection velocity, more explicitly
than we have done here.  For example, the formulae would be expected
to simplify in the limits $\tilde{M}\downarrow 0$ (corresponding to
two solitons differing very much in amplitude) and $\tilde{M}\uparrow
1$ (corresponding to two solitons with nearly the same amplitude).
The calculation of the ejection velocity is challenging because it may require
uniform approximation of $F(q;\tilde{M})$ for all $q$ in the limit
of interest; pointwise asymptotics for fixed $q$ are not enough
to approximate the work integral without further information.

\section{Acknowledgements}
PDM acknowledges the support of NSF grant
DMS 9304580 while a member of the School of Mathematics at the
Institute for Advanced Study.  During the preparation of this paper,
JAB and NNA were affiliated with the Australian Photonics Cooperative
Research Centre.

\vspace{0.4 in}

\noindent{\Large\bf Appendix:  
Formulae for the Two-Particle Force Function Integrand}

\vspace{0.1 in}

\noindent
Here, we record the details of the formulae for the
two-particle force function needed to 
calculate or approximate for special values of 	
$\tilde{M}$ the force and related quantities to any desired 
accuracy.

\vspace{0.1 in}

\noindent{\bf Before averaging.}
The thirteen terms appearing in the sum in (\ref{eq:eqnforh}) are
given here in terms of $c:=\cos\theta$, $S_k:=\sinh(\zeta_k)$ and
$C_k:=\cosh(\zeta_k)$, $\zeta_1:=2m_1(z+y/2)$ and
$\zeta_2:=2m_2(z-y/2)$.  
\begin{displaymath}
h_1=2m_1^6m_2S_2C_1C_2^6 +
2m_1m_2^6S_1C_1^6C_2\hspace{0.3 in}
h_2=
-\Big(m_1^5(m_2^2+m_1^2)S_1C_2^7+m_2^5(m_2^2+m_1^2)
S_2C_1^7\Big)
\end{displaymath}
\begin{displaymath}
h_3=
2m_1^5(m_1^2 + m_2^2c^2)S_1C_2^5 +
2m_2^5(m_1^2c^2+m_2^2)S_2C_1^5
\end{displaymath}
\begin{displaymath}
h_4=-\Big(
2m_1^5(m_2^2+m_1^2)cS_2C_2^5+
2m_2^5(m_2^2+m_1^2)cS_1C_1^5\Big)
\end{displaymath}
\begin{displaymath}
h_5=
m_1^5(m_1^2-9m_2^2)cS_2C_1^2C_2^5 +
m_2^5(m_2^2-9m_1^2)cS_1C_1^5C_2^2
\end{displaymath}
\begin{displaymath}
h_6=
m_1^4m_2(5m_2^2+3m_1^2)cS_1C_1C_2^6+
m_1m_2^4(3m_2^2+5m_1^2)cS_2C_1^6C_2
\end{displaymath}
\begin{displaymath}
h_7= -\Big(
2m_1^4m_2((5m_1^2+m_2^2)c + 4m_2^2c^3)S_1C_1C_2^4 +
2m_1m_2^4((m_1^2+5m_2^2)c + 4m_1^2c^3)S_2C_1^4C_2\Big)
\end{displaymath}
\begin{displaymath}
h_8=
2m_1^4m_2(m_2^2+(5m_1^2+4m_2^2)c^2)S_2C_1C_2^4+
2m_1m_2^4(m_1^2+(5m_2^2+4m_1^2)c^2)S_1C_1^4C_2
\end{displaymath}
\begin{displaymath}
h_9= -\Big(
2m_1^3m_2^4(1+4c^2)S_1C_1^2C_2^5+
2m_1^4m_2^3(1+4c^2)S_2C_1^5C_2^2
\Big)
\end{displaymath}
\begin{displaymath}
h_{10}=
m_1^4m_2(3m_2^2-m_1^2)(1+4c^2)S_2C_1^3C_2^4+
m_1m_2^4(3m_1^2-m_2^2)(1+4c^2)S_1C_1^4C_2^3
\end{displaymath}
\begin{displaymath}
h_{11}= -\Big(
2m_1^3m_2^2(m_2^2-m_1^2)(3c+2c^3)S_2C_1^4C_2^3 +
2m_1^2m_2^3(m_1^2-m_2^2)(3c+2c^3)S_1C_1^3C_2^4\Big)
\end{displaymath}
\begin{displaymath}
h_{12}=
-\Big(
4m_1^2m_2^3((3m_1^2+m_2^2)c+(2m_1^2+4m_2^2)c^3)S_1C_1^3C_2^2+
4m_1^3m_2^2((3m_2^2+m_1^2)c+(2m_2^2+4m_1^2)c^3)S_2C_1^2C_2^3\Big)
\end{displaymath}
\begin{displaymath}
h_{13}=
4m_1^2m_2^3(m_2^2 + (3m_1^2+4m_2^2)c^2 + 2m_1^2c^4)S_2C_1^3C_2^2+
4m_1^3m_2^2(m_1^2+(3m_2^2+4m_1^2)c^2+2m_2^2c^4)S_1C_1^2C_2^3
\end{displaymath}

\vspace{0.1 in}

\noindent{\bf After normalization and averaging.}
Here, we give the nonzero quantities $g_{mn}=g_{mn}(\alpha,\beta)$
appearing in (\ref{eq:eqnforH}).
In these expressions $\beta$ and $\alpha$ are
linked by the normalization condition
$\alpha\beta=1$.  
\begin{displaymath}
\begin{array}{rclrcl}
g_{03}&=&672\alpha^3-672\alpha^7 \hspace{0.4 in}& 
g_{04}&=&1344\alpha^3+3136\alpha^7\\
g_{05}&=&-2304\alpha^3-2816\alpha^7 \hspace{0.4 in}&
g_{06}&=&512\alpha^3+512\alpha^7
\end{array}
\end{displaymath}
\begin{displaymath}
\begin{array}{rcl}
g_{12}&=&-6048\alpha^3-4032\alpha^7+10080\beta\\
g_{13}&=&69664\alpha^3+16576\alpha^7+
1120\alpha^{11}+2240\beta\\
g_{14}&=&
-130544\alpha^3-52016\alpha^7-2960\alpha^{11}-29520\beta\\
g_{15}&=&
75904\alpha^3+48256\alpha^7+4480\alpha^{11}+18816\beta\\
g_{16}&=&
-10368\alpha^3-10368\alpha^7-1920\alpha^{11}-1920\beta
\end{array}
\end{displaymath}
\begin{displaymath}
\begin{array}{rcl}
g_{21}&=&-13440\alpha^3+3360\beta+10080\beta^5\\
g_{22}&=&17920\alpha^3+17920\alpha^7+51520\beta-24640\beta^5\\
g_{23}&=&-234720\alpha^3-48320\alpha^7-5440\alpha^{11}-272640\beta-7200\beta^5\\
g_{24}&=&479872\alpha^3+146176\alpha^7+13184\alpha^{11}+320\alpha^{15}
+352704\beta+36864\beta^5\\
g_{25}&=&-277152\alpha^3-144096\alpha^7-19280\alpha^{11}-560\alpha^{15}
-141808\beta-15120\beta^5\\
g_{26}&=&31680\alpha^3+31680\alpha^7+8800\alpha^{11}+480\alpha^{15}+8800\beta+480\beta^5
\end{array}
\end{displaymath}
\begin{displaymath}
\begin{array}{rcl}
g_{30}&=&-4032\beta+3360\beta^5+672\beta^9\\
g_{31}&=&5600\alpha^3-90944\beta-32480\beta^5-7616\beta^9\\
g_{32}&=&131600\alpha^3-6720\alpha^7+166960\beta+14960\beta^5+15760\beta^9\\
g_{33}&=&-24576\alpha^3-15168\alpha^7+3648\alpha^{11}
+249088\beta+145344\beta^5-1984\beta^9\\
g_{34}&=&-375744\alpha^3-63840\alpha^7-6688\alpha^{11}-480\alpha^{15}
-545056\beta-186144\beta^5 - 9888\beta^9\\
g_{35}&=&280368\alpha^3+108768\alpha^7+13712\alpha^{11}+864\alpha^{15}+16\alpha^{19}
\\&&\hspace{0.3 in}+\,227744\beta+54604\beta^5+2592\beta^9\\
g_{36}&=&-24024\alpha^3-24024\alpha^7-8008\alpha^{11}-728\alpha^{15}-8\alpha^{19}
-8008\beta-728\beta^5-8\beta^9
\end{array}
\end{displaymath}
\begin{displaymath}
\begin{array}{rcl}
g_{40}&=&-6720\beta-22400\beta^5-2240\beta^9\\
g_{41}&=&31680\alpha^3+216128\beta+162400\beta^5+19072\beta^9+800\beta^{13}\\
g_{42}&=&-269952\alpha^3-18720\alpha^7
-592576\beta-268928\beta^5-29984\beta^9-2560\beta^{13}\\
g_{43}&=&384912\alpha^3+82032\alpha^7+2976\alpha^{11}
+387040\beta+43584\beta^5-8784\beta^9+1168\beta^{13}\\
g_{44}&=&-81312\alpha^3-63840\alpha^7-8256\alpha^{11}-192\alpha^{15}\\
&&\hspace{0.3 in}+\,97152\beta+127104\beta^5+26784\beta^9+864\beta^{13}\\
g_{45}&=&-61776\alpha^3+4368\alpha^{11}+288\alpha^{15}
-96096\beta-39312\beta^5-4032\beta^9-48\beta^{13}
\end{array}
\end{displaymath}
\begin{displaymath}
\begin{array}{rcl}
g_{50}&=&9216\beta+21760\beta^5+4864\beta^9\\
g_{51}&=&-20096\alpha^3-129152\beta-155264\beta^5-38272\beta^9-1280\beta^{13}\\
g_{52}&=&112848\alpha^3+7584\alpha^7
+355920\beta+307872\beta^5+71872\beta^9+3984\beta^{13}+80\beta^{17}\\
g_{53}&=&-165584\alpha^3-22752\alpha^7-528\alpha^{11}\\
&&\hspace{0.3 in}-\,330528\beta-200112\beta^5-32672\beta^9-1392\beta^{13}-96\beta^{17}\\
g_{54}&=&72072\alpha^3+15288\alpha^7+648\alpha^{11}\\
&&\hspace{0.3 in}+\,92664\beta+24024\beta^5-6552\beta^9-1512\beta^{13}-24\beta^{17}
\end{array}
\end{displaymath}
\begin{displaymath}
\begin{array}{rcl}
g_{60}&=&-1536\beta-4096\beta^5-1536\beta^9\\
g_{61}&=&1280\alpha^3+13568\beta+27648\beta^5+13568\beta^9+1280\beta^{13}\\
g_{62}&=&-4096\alpha^3-96\alpha^7
-27808\beta-50688\beta^5-27808\beta^9-4096\beta^{13}-96\beta^{17}\\
g_{63}&=&2912\alpha^3+112\alpha^7
+16016\beta+27456\beta^5+16016\beta^9+2912\beta^{13}+112\beta^{17}
\end{array}
\end{displaymath}

\end{document}